\documentclass[fleqn,usenatbib]{mnras}
\usepackage{newtxtext,newtxmath}
\usepackage[T1]{fontenc}
\usepackage{ae,aecompl}
\usepackage{color,soul}

\PassOptionsToPackage{normalem}{ulem}
\usepackage{ulem}

\newcommand{\added}[1]{#1}
\newcommand{\deleted}[1]{}

\usepackage{graphicx}	% Including figure files
\usepackage{amsmath}	% Advanced maths commands
\usepackage{amssymb}	% Extra maths symbols
\usepackage{wasysym}
\newcommand{\angstrom}{\textup{\AA}}
\newcommand{\gcm}{\,g\,cm$^{-3}$}
\title[Diversity of Si~II linewidths in SNe Ia]{An asymmetric explosion mechanism may explain the diversity of Si~II linewidths in Type Ia supernovae}

\author[R. Livneh et al.]{
Ran Livneh,$^{1}$\thanks{E-mail: ran.livneh@weizmann.ac.il}
Boaz Katz$^{1}$
\\
$^{1}$Department of Particle Physics \& Astrophysics, The Weizmann Institute of Science, Rehovot 76100, Israel \\
}

\date{Accepted XXX. Received YYY; in original form ZZZ}

\pubyear{2020}

\begin{document}
\label{firstpage}
\pagerange{\pageref{firstpage}--\pageref{lastpage}}
\maketitle

%%%%%%%%%%%%%%%%%%%%%%%%%%%%%%%%%%%%%%%%%%%%%%%%%%
%%%%%%%%%%%%%%%%%%% Abstract %%%%%%%%%%%%%%%%%%%%%

\begin{abstract}
Near maximum brightness, the spectra of \deleted{t}\added{T}ype Ia supernovae (SNe Ia) present typical absorption features of Silicon II observed at roughly $6100\angstrom$ and $5750\angstrom$. The 2-D distribution of the pseudo-equivalent widths (pEWs) of these features is a useful tool for classifying SNe Ia spectra (Branch plot). 
Comparing the observed distribution of SNe on the Branch plot to results of simulated explosion models, we find that 1-D models fail to cover most of the distribution. 
In contrast, we find that \textsc{Tardis} radiative transfer simulations of the WD head-on collision models along different lines of sight almost fully cover the distribution. 
We use several simplified approaches to explain this result. We perform order-of-magnitude analysis and model the opacity of the \ion{Si}{II} lines using LTE and NLTE approximations.
Introducing a simple toy model of spectral feature formation, we show that the pEW is a good tracer for the extent of the absorption region in the ejecta. 
Using radiative transfer simulations of synthetic SNe ejecta, we reproduce the observed Branch plot distribution by varying the luminosity of the SN and the Si density profile of the ejecta. 
We deduce that the success of the collision model in covering the Branch plot is a result of its asymmetry, which allows for a significant range of Si density profiles along different viewing angles, uncorrelated with a range of $^{56}$Ni yields that cover the observed range of SNe Ia luminosity. 
We use our results to explain the shape and boundaries of the Branch plot distribution.
\end{abstract}


%%%%%%%%%%%%%%%%%%%%%%%%%%%%%%%%%%%%%%%%%%%%%%%%%%
%%%%%%%%%%%%%%%%% BODY OF PAPER %%%%%%%%%%%%%%%%%%

\section{Introduction}

%%%%%%%%%%%%%%%%%%%%%%%%%%%%%%%%%%%%%%%%%%%%%%%%%%

\begin{figure*}
    \centering
    \includegraphics[width=\columnwidth]{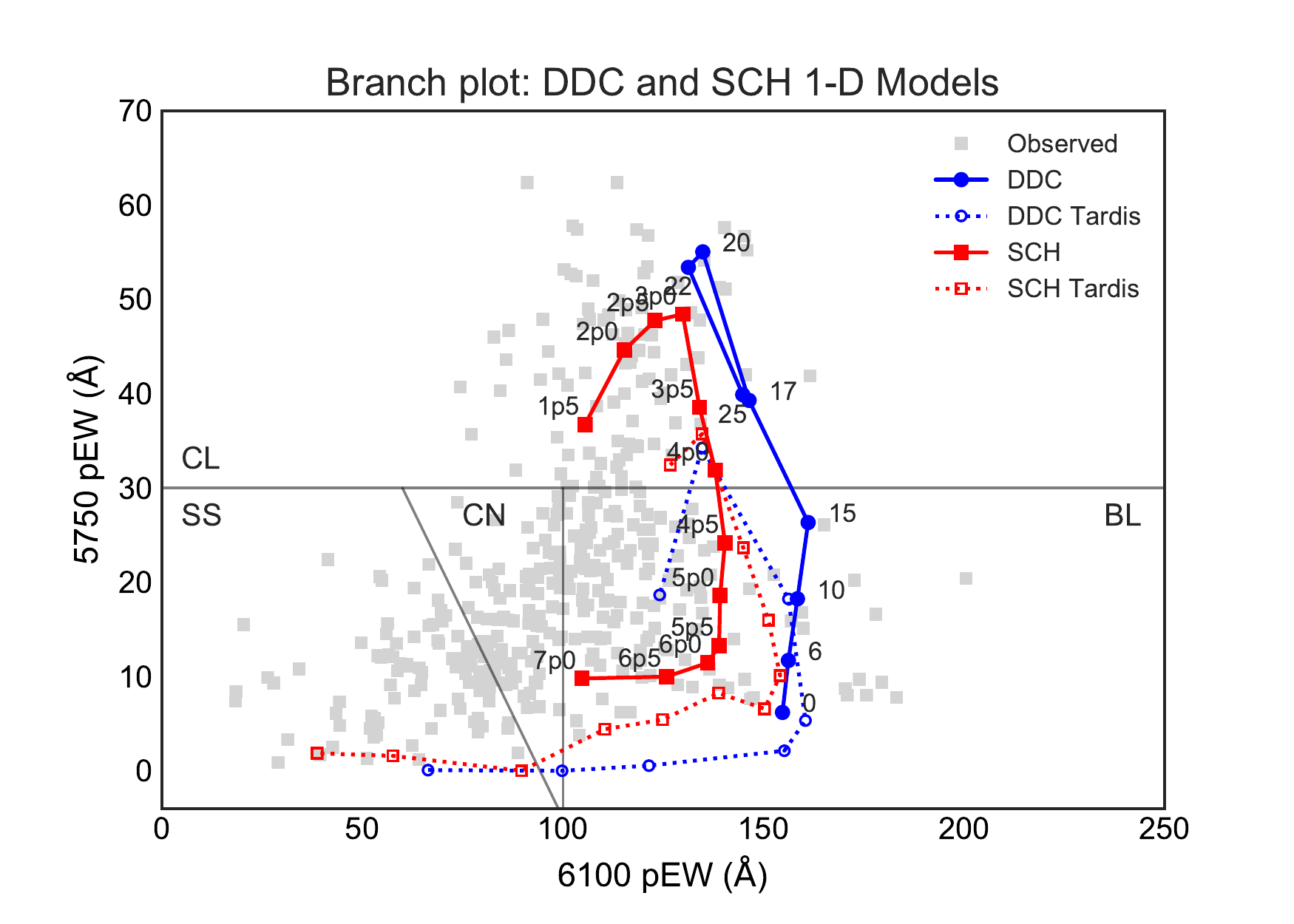}
    ~
    \includegraphics[width=\columnwidth]{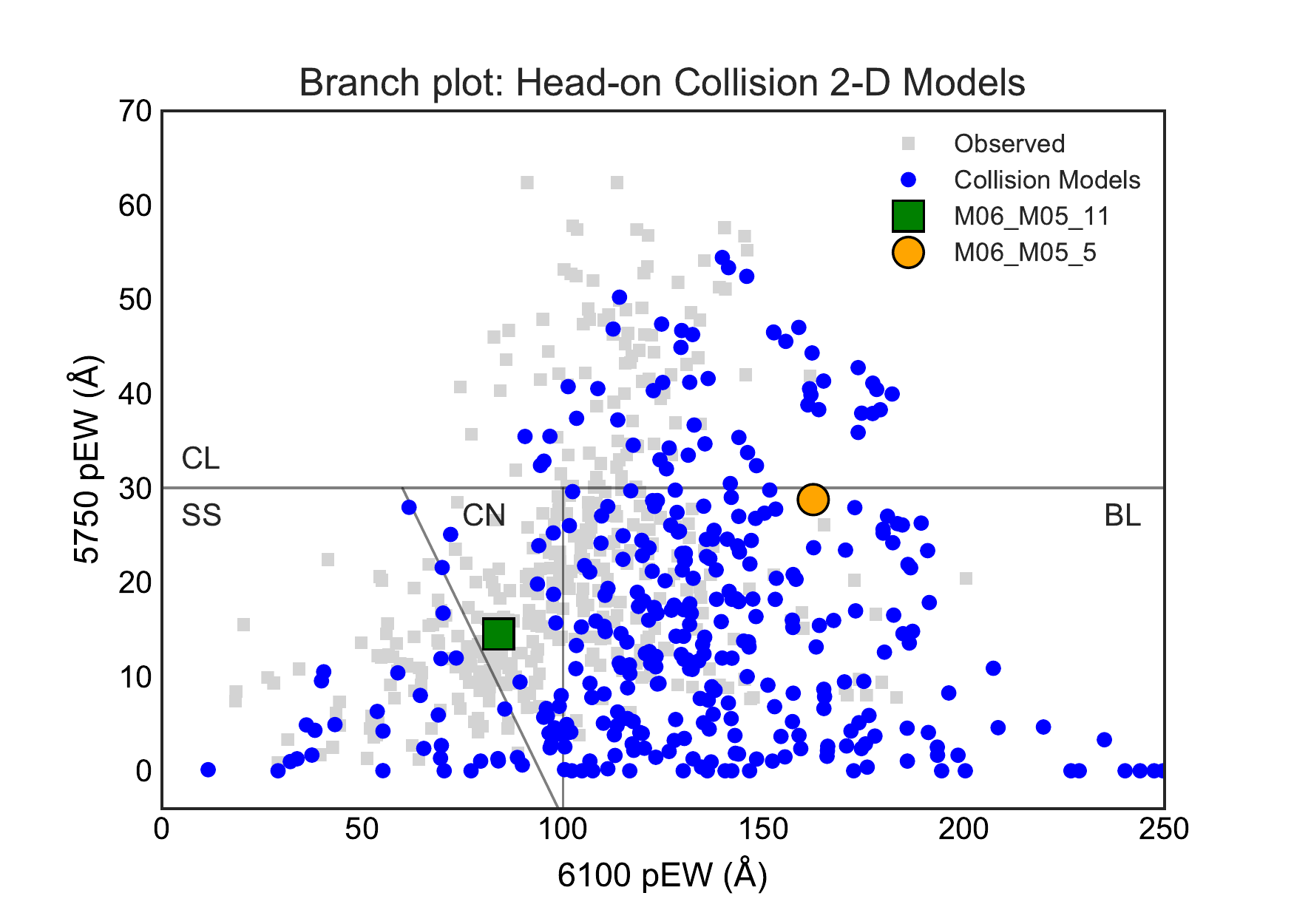}
      
    \caption{Branch plot of simulated models overlaid on observed CfA  \citep{CfA2012}, CSP \citep{CSP2013} and BSNIP \citep{Silverman2012} data ($\pm5$ days from peak). Borders between Branch types follow \citet{Silverman2012}. \textbf{Left:} Solid blue and red lines represent 1-D delayed detonation (DDC) and sub-Chandrasekhar models (SCH, see \S\ref{sec:models}). The pEWs were extracted from numerically calculated spectra from \citet{Blondin2013,Blondin2017}\deleted{ - m}\added{. M}odel parameters can be found in Table \ref{table:models}. Dotted lines are the same models with radiative transfer simulated using \textsc{Tardis} (this work). \textbf{Right:} Head-on direct collision models \citep{Kushnir2013} with radiative transfer simulated using \textsc{Tardis} for different viewing angles (\S\ref{sec:collision}). Two viewing angles of the same collision event between a $0.6\,\textup{M}_\odot$ and a $0.5\,\textup{M}_\odot$ WD with $M(^{56}$Ni$)=0.27\,\textup{M}_\odot$ are highlighted in green and orange (see Fig.~\ref{fig:slices}). Corresponding Si density profiles are shown in the same colors in Fig.~\ref{fig:Density_Models}.}

%    \caption{Branch plot of simulated models overlaid on observed CfA  \citep{CfA2012}, CSP \citep{CSP2013} and BSNIP \citep{Silverman2012} data ($\pm5$ days from peak). Borders between Branch types follow \citet{Silverman2012}. \textbf{Left:} Solid blue and red lines represent \deleted{1-D delayed detonation (DDC) and sub-Chandrasekhar }\added{DDC and SCH }models (\deleted{SCH, }see \S\ref{sec:models}). The pEWs were extracted from numerically calculated spectra from \citet{Blondin2013,Blondin2017}\deleted{ - m}\added{. M}odel parameters can be found in Table \ref{table:models}. Dotted lines are the same models with radiative transfer simulated using \textsc{Tardis} (this work). \textbf{Right:} Head-on direct collision models \citep{Kushnir2013} with radiative transfer simulated using \textsc{Tardis} for different viewing angles (\S\ref{sec:collision}). Two viewing angles of the same collision event between a $0.6\,\textup{M}_\odot$ and a $0.5\,\textup{M}_\odot$ WD with $M(^{56}$Ni$)=0.27\,\textup{M}_\odot$ are highlighted in green and orange (see Fig.~\ref{fig:slices}). Corresponding Si density profiles are shown in the same colors in Fig.~\ref{fig:Density_Models}.}

    \label{fig:Branch2}
\end{figure*}

%%%%%%%%%%%%%%%%%%%%%%%%%%%%%%%%%%%%%%%%%%%%%%%%%%

\begin{figure}
	\includegraphics[width=\columnwidth]{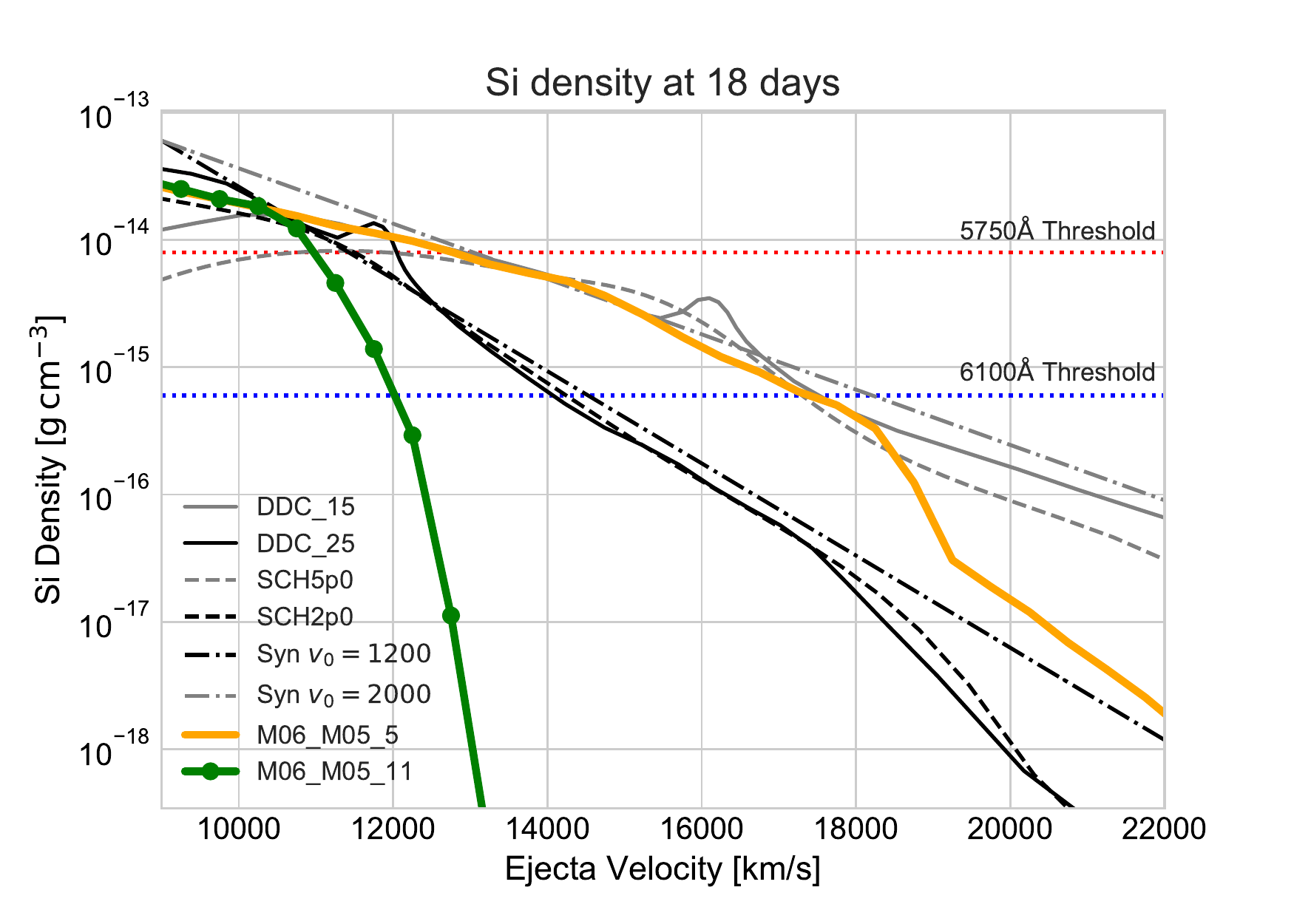}

    \caption{Si density at 18 days, as a function of velocity within the ejecta for various models: delayed detonations (DDC), sub-Chandrasekhar (SCH) \deleted{\mbox{\citep{Blondin2013,Blondin2017}}}\added{(see \S\mbox{\ref{sec:models}})}, synthetic exponential (\texttt{Syn}, see \S\ref{sec:synthetic}) and head-on collisions (\S\ref{sec:collision}). Two examples are shown for each model. For the head-on collision model, two viewing angles of the same collision are shown (angle no. 5 and 11), in correspondence with Fig.~\ref{fig:Branch2} and Fig.~\ref{fig:slices}. In \texttt{M06\_M05\_5} the Si is extended whereas in \texttt{M06\_M05\_11} the Si density drops steeply. The blue and red dotted lines are approximate Si density thresholds above which $\tau_s>1$ for the $6100\angstrom$ and $5750\angstrom$ features, derived in \S\ref{sec:threshold}.}
    \label{fig:Density_Models}
\end{figure}

%%%%%%%%%%%%%%%%%%%%%%%%%%%%%%%%%%%%%%%%%%%%%%%%%%

There is strong evidence that \deleted{t}\added{T}ype Ia supernovae (SNe Ia) are the product of thermonuclear explosions of white dwarfs (WDs), yet the nature of the progenitor systems and the mechanism that triggers the explosion remain long-standing open questions (see e.g. \citealt{Maoz2014,Livio2018,Soker2019} for recent reviews). Optical spectra at the photospheric phase are a sensitive probe of the structure and composition of SNe Ia ejecta. The P-Cygni absorption lines (superimposed upon a pseudo-continuum) are Doppler shifted and widened and their shape is directly related to the velocity distribution of the absorbing ions. The observed spectra show significant diversity in line depths \citep[e.g.][]{Branch2006}, shifts \citep[e.g.][]{Wang2009} and time evolution \citep[e.g.][]{Benetti2005} which is partly correlated with the luminosity that covers a range of about one order of magnitude. Whether the observed diversity is a result of multiple explosion mechanisms or due to a continuous range of underlying parameters in a single explosion mechanism is a key question in addressing the Type Ia problem.  

Two especially useful features in the near-peak spectra of SNe Ia are the \ion{Si}{II} features at $5750\angstrom$ and $6100\angstrom$, attributed to \ion{Si}{II} $\lambda 5972$ and \ion{Si}{II} $\lambda 6355$.
The 2-D distribution of the pseudo-equivalent widths (pEW) of these features (see Fig.~\ref{fig:Branch2}) was introduced by \citet{Branch2006}, classifying spectra into four groups: core normal (CN), broad line (BL), cool (CL), and shallow silicon (SS). It was shown that adjacent SNe on the plot exhibit overall similar spectra at maximum light, indicating that the variation in the two pEWs captures most of the observed diversity in near-peak spectra. The $6100\angstrom$ feature spans a large range of pEWs ($20\angstrom$ \deleted{-}\added{to} $200\angstrom$) while the $5750\angstrom$ feature spans a smaller range ($0$ \deleted{-}\added{to} $70 \angstrom$) and the two pEWs are not correlated (though the range of $6100\angstrom$ pEWs decreases with increasing pEW of the $5750\angstrom$ feature). 
The span in observed pEWs is affected by the presence of Si at varying velocities and by the properties of the radiation field and electron density that determine the ionization and excitation level of the ions.

In order to relate the spectral features to the structure of explosion models, it would be useful to relate them to the distribution of Si in the ejecta. However, the fraction of Si in the relevant ionization (single) and excitation levels is of the order of $10^{-9}$ (see \deleted{Fig.~3, }\S\ref{sec:properties}), and the optical depth depends on the properties of the plasma. In fact, the ratio of the depth of the two features is correlated with brightness in a continuous way \citep{Nugent1995} and is understood to be set by the temperature which is largely determined by the luminosity \citep[e.g.][]{Hachinger2008}. Using the photospheric spectral synthesis code \textsc{Tardis} \citep{Tardis}, it was recently shown by \citet{Heringer2017} that sequences of ejecta with the same structure and composition but with varying luminosities can (approximately) continuously connect the spectra of bright and faint \deleted{t}\added{T}ype Ia's. While these results demonstrate the role of the variations in the radiation field and strengthen the case for a single underlying mechanism, it is clear that the distribution of spectral features does not constitute a one-parameter family set by luminosity alone \citep[e.g.][]{Hatano2000}. 

As an illustration, the results of two\added{ main} classes of spherically symmetric models that span the entire range of luminosities of \deleted{t}\added{T}ype Ia's are shown in the left panel of Fig.~\ref{fig:Branch2} based on the results of \citet{Blondin2013,Blondin2017} \deleted{-}\added{--} central detonations of sub-Chandrasekhar WDs (\deleted{SCH, }e.g. \citealt{Sim2010}) and delayed detonation Chandrasekhar models (\deleted{DDC, }e.g. \citealt{Nomoto1982, Khokhlov1991}). As can be seen, while the \ion{Si}{II} line pEWs are in the right ball-park, they tend to the right side of the plot and cannot account for the 2-D distribution of observed pEWs.\added{ An exhaustive comparison of existing 1-D models is beyond the scope of this paper, but see for another example} \deleted{Another example can be found in }Fig.~16 of \citet{Wilk2018}, where various 1-D model results are clustered near the BL region of the plot. Specifically, core normal (CN) SNe Ia are especially challenging to reproduce \citep[e.g.][]{Townsley2019}.

In this paper we extend the study of \citet{Heringer2017} using similar approximations (in particular the \textsc{Tardis} code), but including varying ejecta structures, and accounting for NLTE effects critical for quantitative analysis of the $5750\angstrom$ feature. In particular, we show that a single asymmetric explosion model, namely head-on collisions of WDs (e.g. \mbox{\citealt{Rosswog2009}}, \mbox{\citealt{Raskin2010}}, \mbox{\citealt{Kushnir2013}}), can reach the entire extent of the observed distribution of line pEWs (see right panel of Fig.~\ref{fig:Branch2}). This is due to the significant range of Si density profiles (see Fig.~\ref{fig:Density_Models}), which include profiles with Si extending to $20,000$~km/s, but also profiles with sharp cutoffs at $v\lesssim13,000$~km/s\added{, depending on the viewing angle. The same collision models, when averaged over viewing angle and simulated as 1-D models, show only extended Si profiles and tend to the right side of the Branch plot (not plotted) like the other tested models. The head-on collision model is used here to demonstrate the possible role of asymmetry in reproducing the observed Branch plot}.  

An accurate calculation of the pEWs of the \ion{Si}{II} features requires a self-consistent solution of the radiation transfer problem, coupled to the solution at each location of the ionization balance and the level excitation equilibrium, which deviate from local-thermal-equilibrium (LTE). \textsc{Tardis} adopts crude approximations for the radiation field and ionization balance (while solving the non-LTE excitation equations, see \S\ref{sec:radiative_transfer}). For this reason, our results should be treated as a proof-of-concept rather than as accurate estimates. A rough estimate of the accuracy is obtained by the comparison of our \textsc{Tardis} calculations for the same ejecta as those used by \citet{Blondin2013,Blondin2017}, who solve the radiation transfer problem directly, which is shown in the left panel of Fig.~\ref{fig:Branch2}. As can be seen, while there are significant differences for each ejecta, the sequence is qualitatively similar, with larger differences at higher temperatures (bottom of the plot, see \S\ref{sec:models}). 

The outline of this paper is as follows: In \S\ref{sec:formation} an order-of-magnitude analysis of the formation of the \ion{Si}{II} features is performed. In \S\ref{sec:toy_model}, a toy model of an ejecta containing a single, fully absorbing line is simulated. The resulting spectral features are shown to reproduce the non-trivial relations between the pEW and both the fractional depth and the FWHM. In \S\ref{sec:radiative_transfer} we describe our use of the \textsc{Tardis} radiative transfer simulation and present results for several hydrodynamic models and synthetic ejecta, exploring the dependence of the Si features on ejecta composition and SN luminosity. In \S\ref{sec:threshold} we use our model to numerically find an approximate Si density threshold predicting the extent of the absorption region and the pEW of the $6100\angstrom$ feature. Finally, in \S\ref{sec:boundaries} we use our results to explain the boundaries of the Branch plot. 

The observed SNe sample used in this paper is based on data from the Center for Astrophysics Supernova Program (CfA, \citealt{CfA2012}), the Carnegie Supernova Project (CSP, \citealt{CSP2013}), and the Berkeley Supernova Ia Program (BSNIP, \citealt{Silverman2012}). We also use fractional depths and FWHM data from BSNIP in our study of the behavior of spectral features.

%%%%%%%%%%%%%%%%%%%%%%%%%%%%%%%%%%%%%%%%%%%%%%%%%%
%%%%%%%%%%%%%%%%%%%%%%%%%%%%%%%%%%%%%%%%%%%%%%%%%%

\section{Basic properties of the two \ion{Si}{II} features}
\label{sec:formation}

In this section, we review the basic properties of the $6100\angstrom$ and $5750\angstrom$ features and explore their relation to typical properties of \deleted{t}\added{T}ype Ia supernovae. 

%%%%%%%%%%%%%%%%%%%%%%%%%%%%%%%%%%%%%%%%%%%%%%%%%%

\subsection{Required density of excited \ion{Si}{II} ions for absorption}
\label{sec:properties}

The Sobolev optical depth for interaction is set by the local number density $n_l$ of the ions in the lower excitation state of the transition. The required number density and mass density of Si ions in the lower excited level for a Sobolev optical depth of unity at a time $t_{\rm exp}= 18~t_{18\rm d}~\rm days$ after explosion and transition wavelength of $\lambda \approx 6000~\angstrom$ are about:
%%%%%%%%%%%%%%%%%%%%%%%%%%%%%%%%%%%%%%%%%%%%%%%%%%
\begin{align}
n_{l,\min}&=\frac{1}{\lambda r_e t_{\rm exp}~\textup{c}\pi f}\sim 0.4~ f^{-1}t_{18\rm d}^{-1}~\rm cm^{-3}\cr
\rho_{l,\min}&=28 \textup{m}_{\rm p} n_{l,\min}\sim  2\times10^{-23} f^{-1}t_{18\rm d}^{-1}~\rm g~cm^{-3}
\label{eq:sobolev}
\end{align}
%%%%%%%%%%%%%%%%%%%%%%%%%%%%%%%%%%%%%%%%%%%%%%%%%%
where $f$ is the oscillator strength of the transition (see Table~\ref{table:osc_strengths}), $r_e=\rm e^2/(m_ec^2)$ is the classical radius of the electron and the correction for stimulated emission is ignored. 

The required density of excited ions is smaller by orders of magnitude than the typical density of a \deleted{t}\added{T}ype Ia expanding at $v=10^9 v_9~ \rm cm/s$ at similar epochs (see Fig.~\ref{fig:Density_Models}),
%%%%%%%%%%%%%%%%%%%%%%%%%%%%%%%%%%%%%%%%%%%%%%%%%%
\begin{equation}
\rho\sim \frac{\textup{M}_{\odot}}{\frac{4\pi}{3}(vt_{\rm exp})^3}\sim 10^{-13}{v}_9^{-3}t_{18\rm d}^{-3}~\rm g~cm^{-3}.
\end{equation}
%%%%%%%%%%%%%%%%%%%%%%%%%%%%%%%%%%%%%%%%%%%%%%%%%%
This does not always result in high optical depth however, since only a very small fraction of the Si is in the required ionization and excitation states as demonstrated below. 

\subsection{Rough estimates using the LTE approximation}
\label{sec:properties2}

\begin{figure}
	\includegraphics[width=\columnwidth]{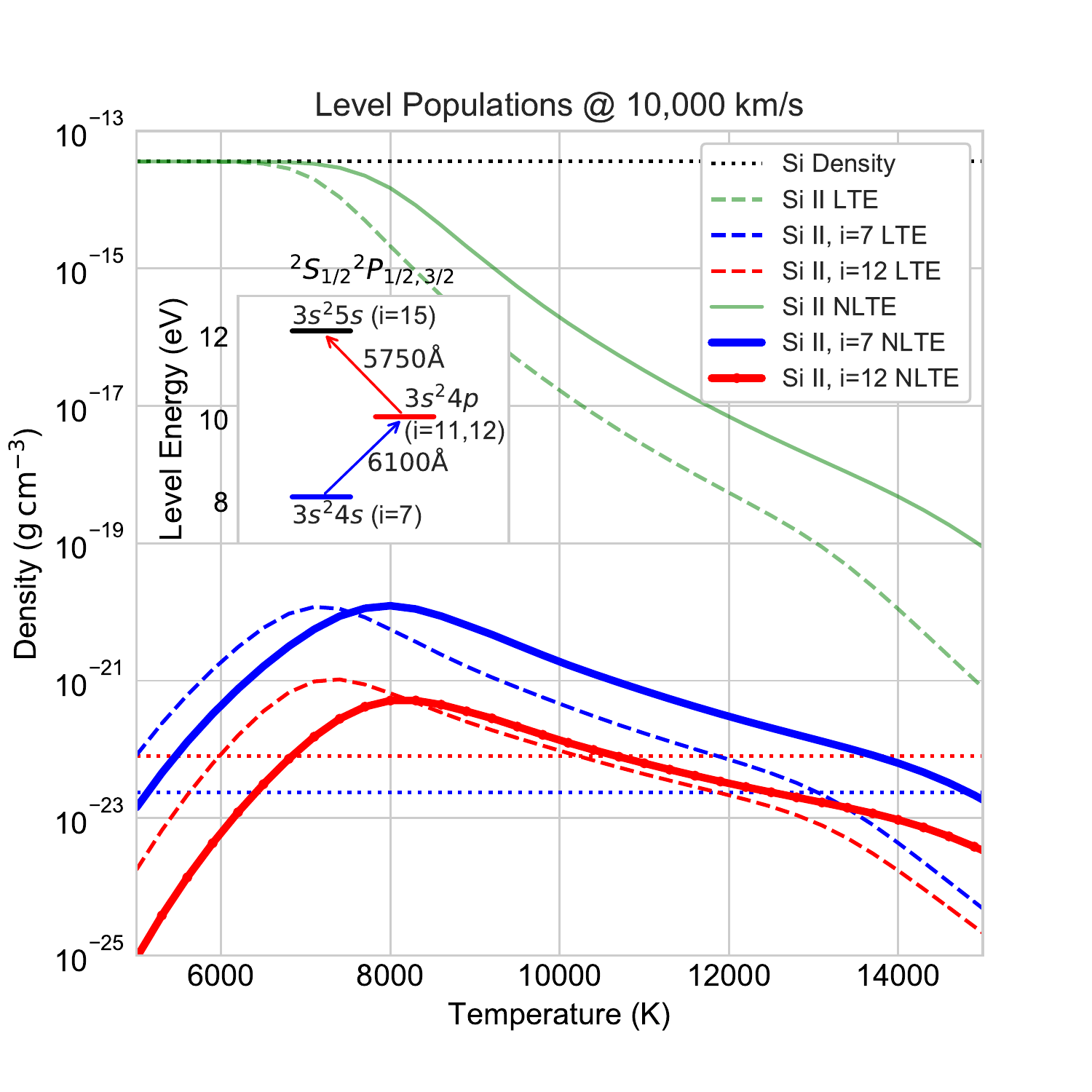}
	\caption{Density of \ion{Si}{II} ions excited to the levels relevant for the $6100\angstrom$ and $5750\angstrom$ lines ($i=7,12$ \deleted{-}\added{--} only one line from each doublet is shown) as a function of temperature\deleted{,}\added{.} \deleted{taking a typical density}\added{Total density is $\rho=6\times10^{-14}$\gcm} at $10,000$~km/s (\deleted{$\rho=6\times10^{-14}$\gcm, }black dotted line) and \deleted{a typical}\added{the} elemental abundance\added{ is taken to be} \deleted{(}$60\%$ Si, $30\%$ S, $5\%$ Ca, $5\%$ Ar\deleted{)}. Green lines represent total \ion{Si}{II} density, blue lines represent the density of \ion{Si}{II} ions excited to the lower level of the $6100\angstrom$ transition and red lines are the same for the $5750\angstrom$ transition. Dashed lines are at LTE; solid lines assume \texttt{nebular} ionization with a dilution factor of $W=0.3$ and are based on a detailed calculation of NLTE excitation equilibrium. Dotted blue and red lines represent the necessary $n_l$ for $\tau_s=1$ at 18 days without stimulated emission. The inset summarizes the $6100\angstrom$ and $5750\angstrom$ transitions.}
	\label{fig:Saha}
\end{figure}

A zeroth-order estimate of the fraction of Si in the correct ionization and excitation level can be obtained by assuming LTE, solving the Saha equation for the ionization and finding the excitation fractions from the Boltzmann distribution. The $6100\angstrom$ and $5750\angstrom$ features arise from the (doublet) transitions $3s^24s\rightarrow 3s^24p$ with rest-frame wavelength $6355 \angstrom$ and $3s^24p\rightarrow 3s^25s$ with rest-frame wavelength $5972 \angstrom$, respectively, which are blue-shifted by about $10,000~\rm km/s$ (see inset of Fig.~\ref{fig:Saha} and Table~\ref{table:osc_strengths}). 
The density of \ion{Si}{II} in the lower excitation levels for the two transitions assuming LTE is shown in dashed lines in Fig.~\ref{fig:Saha} as a function of temperature for an adopted total density of $\rho=6\times10^{-14}$\gcm~and composition by mass of 60\% Si, 30\% S, 5\% Ca, 5\% Ar \added{\mbox{\citep[e.g. similar to][]{Nomoto1984}}} (elements other than Si having a weak effect on the free electron density and ionization). 
The required density for obtaining a Sobolev optical depth equal to unity at 18 days for the two features based on equations \eqref{eq:sobolev} is shown in Fig.~\ref{fig:Saha} in blue and red dotted lines. 
%%%%%%%%%%%%%%%%%%%%%%%%%%%%%%%%%%%%%%%%%%%%%%%%%%
\begin{table}
	\centering
	\begin{tabular}{ |c|c|c|c|c| } 
		\hline
		Wavelength ($\angstrom$) & Transition & $f$ & $g_l$ & $E_l~(\rm eV)$ \\
		\hline
		$6371$ & $7\rightarrow11$  & 0.414 & 2 & 8.121 \\ 
		$6347$ & $7\rightarrow12$  & 0.705 & 2 & 8.121 \\ 
		$5957$ & $11\rightarrow15$ & 0.298 & 2 & 10.066 \\ 
		$5978$ & $12\rightarrow15$ & 0.303 & 4 & 10.073 \\ 
		\hline
	\end{tabular}
	\caption{Approximate parameters for the two \ion{Si}{II} doublets (see inset, Fig.~\ref{fig:Saha}). Data from NIST \citep{NIST_ASD}.}
	\label{table:osc_strengths}
\end{table}
%%%%%%%%%%%%%%%%%%%%%%%%%%%%%%%%%%%%%%%%%%%%%%%%%%

%%%%%%%%%%%%%%%%%%%%%%%%%%%%%%%%%%%%%%%%%%%%%%%%%%

%%%%%%%%%%%%%%%%%%%%%%%%%%%%%%%%%%%%%%%%%%%%%%%%%%

The photospheric temperature for typical luminosities of $L=10^{43} L_{43}~\rm erg~s^{-1}$ is of order:
%%%%%%%%%%%%%%%%%%%%%%%%%%%%%%%%%%%%%%%%%%%%%%%%%%
\begin{equation}
T_{\rm eff}=\left(\frac{L/r_{\rm ref}}{4\pi (vt_{\rm exp})^2\sigma_{\rm B}}\right)^{1/4}\sim 10400~L_{43}^{1/4}{v_9}^{-1/2}t_{18\rm d}^{-1/2}r_{0.5}^{-1/4}~\rm K
\end{equation}
%%%%%%%%%%%%%%%%%%%%%%%%%%%%%%%%%%%%%%%%%%%%%%%%%%
where $r_{\rm ref}=0.5~r_{0.5}$ is the (inverse) suppression of the luminosity compared to that of a free surface due to the reflection of many of the photons back to the photosphere. 
As can be seen in the figure, the typical density of ions for the relevant temperature of $10,000\rm\,K$ is about 10 \deleted{-}\added{to} 100 times the required density for optical depth of unity. Given that the Si density naturally drops by more than 2 orders of magnitudes between $10,000~\rm km/s$ and $20,000~\rm km/s$, it is reasonable that \deleted{t}\added{T}ype Ia's have absorption regions that are within this range.

As can be seen in Fig.~\ref{fig:Saha}, at temperatures below about $7000\rm\,K$, all Si atoms are singly ionized (\ion{Si}{II}, green dashed line). For higher temperatures, the abundance of \ion{Si}{II} diminishes quickly, giving way to doubly ionized \deleted{\mbox{\ion{Si}{II}I}}\added{\mbox{\ion{Si}{III}}}. An opposite effect occurs for the excitation \deleted{-}\added{--} as the temperature rises, the higher excited levels which are necessary for the $5750\angstrom$ and $6100\angstrom$ transitions are increasingly populated. The combined effect of ionization and excitation is that for cooler ejecta (down to $\sim7000\rm\,K$ at $V_{\rm ph}$), the optical depth $\tau_s$ at the photosphere of both lines is larger (see also \citealt{Hachinger2008}). 
As one moves out through the ejecta, the density drops, causing the optical depth to decrease, finally dropping below unity at a critical velocity affected by the initial value at the photosphere\deleted{ - f}\added{. F}or lower temperatures at the photosphere, the extent of the absorption region is increased, resulting in a larger pEW.
Below a critical temperature of $\sim7000\rm\,K$ (for LTE) we identify a saturation effect: the population of both of the relevant excited levels peaks and drops for lower temperatures. The implications of this effect will be discussed in \S\ref{sec:low_lumin}.

%%%%%%%%%%%%%%%%%%%%%%%%%%%%%%%%%%%%%%%%%%%%%%%%%%
%%%%%%%%%%%%%%%%%%%%%%%%%%%%%%%%%%%%%%%%%%%%%%%%%%

\section{Toy model for spectral features}
\label{sec:toy_model}

%%%%%%%%%%%%%%%%%%%%%%%%%%%%%%%%%%%%%%%%%%%%%%%%%%

\subsection{A simple absorption model reproduces relations between line parameters}
\label{sec:extent}

\begin{figure*}
	\centering
	\includegraphics[width=\columnwidth]{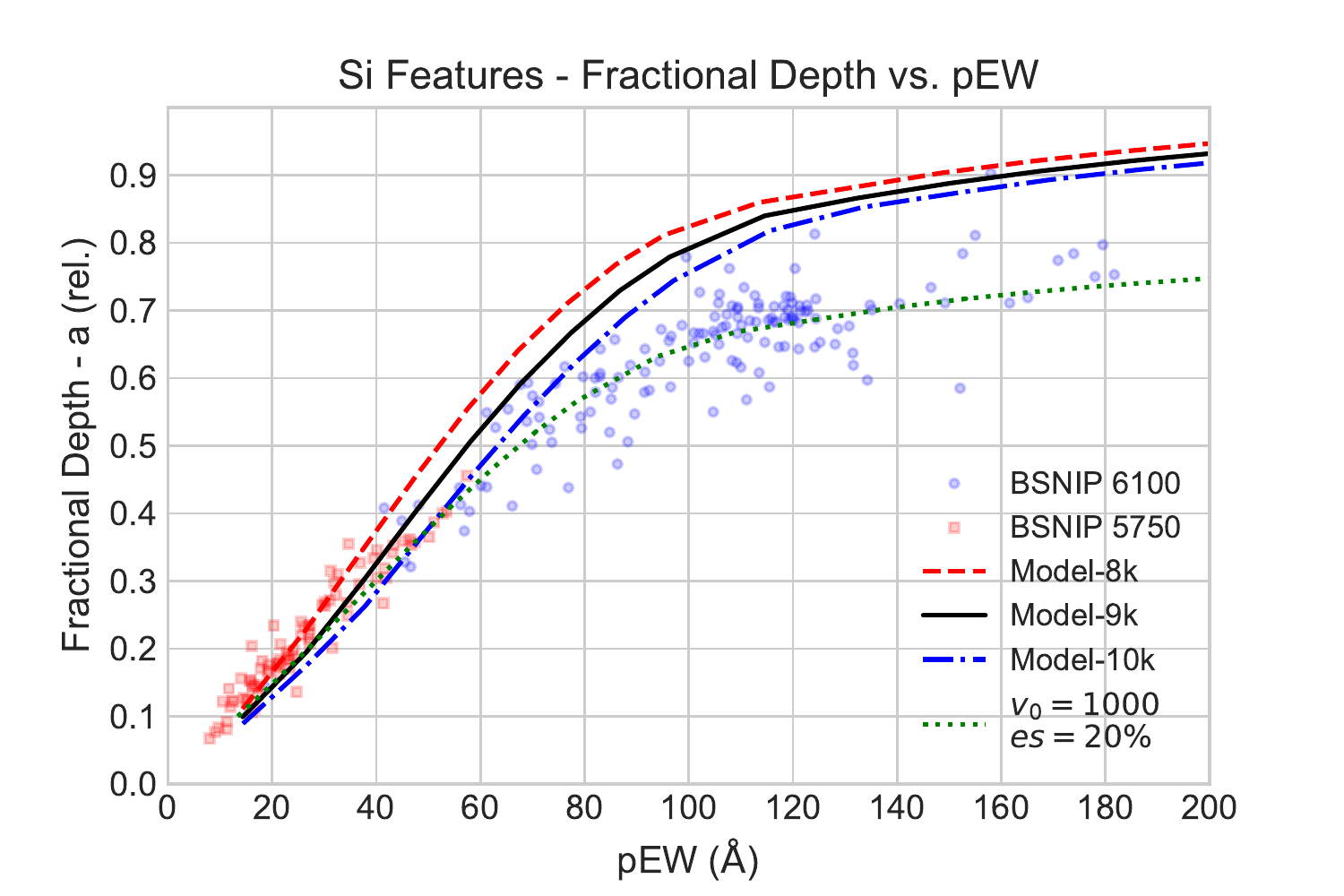}
	~
	\includegraphics[width=\columnwidth]{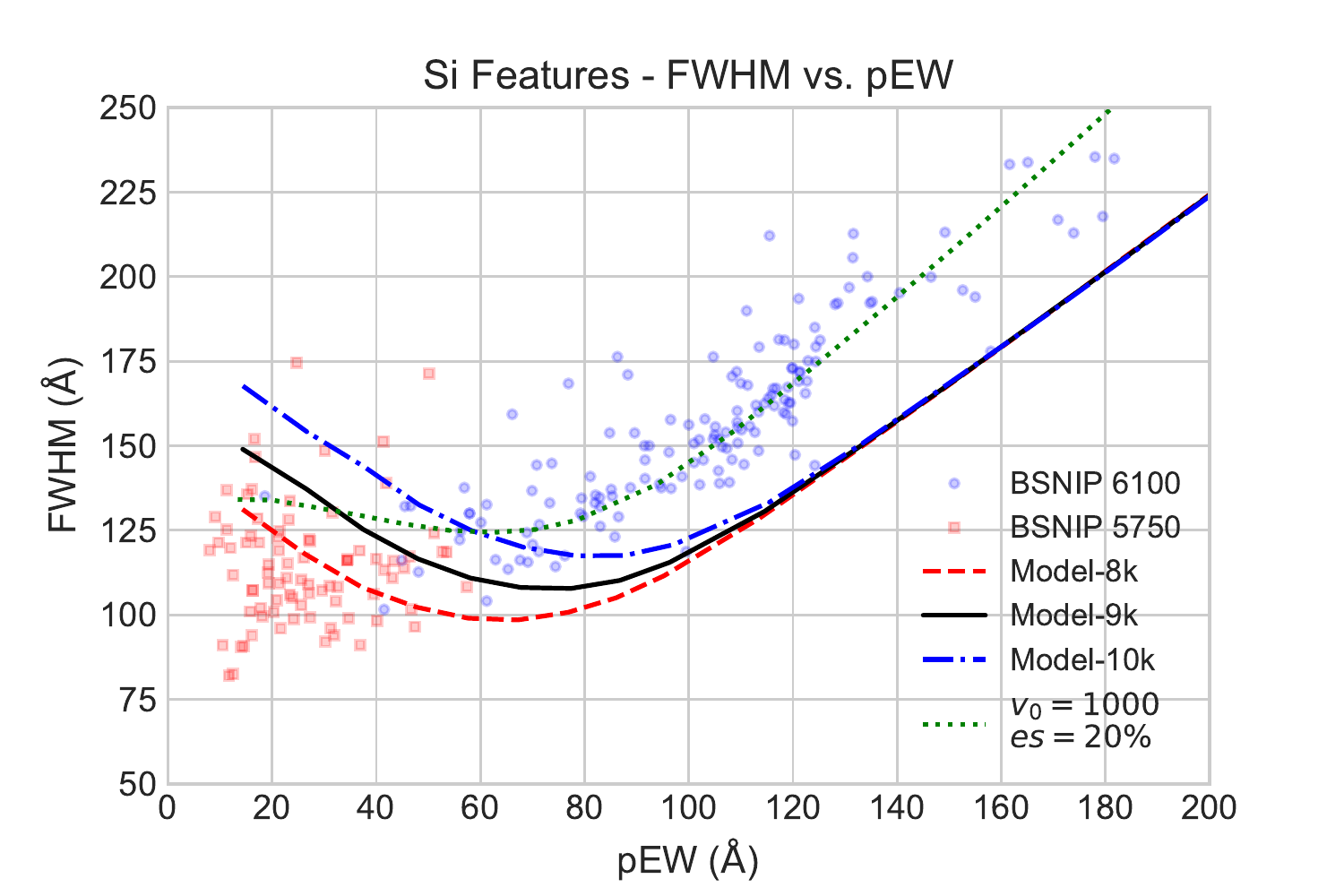}
	\hfill
	\includegraphics[width=\columnwidth]{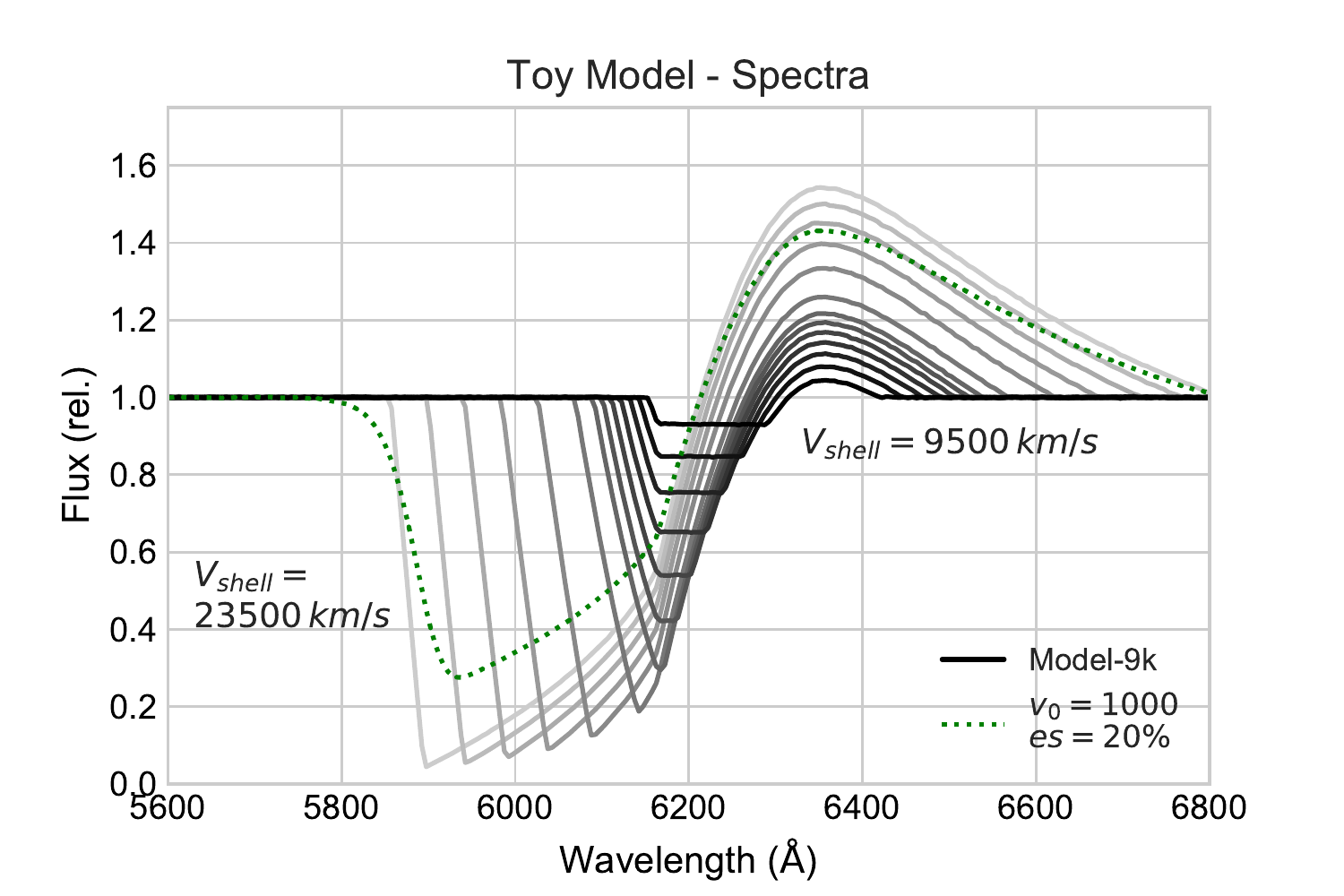}
	~ 
	\includegraphics[width=\columnwidth]{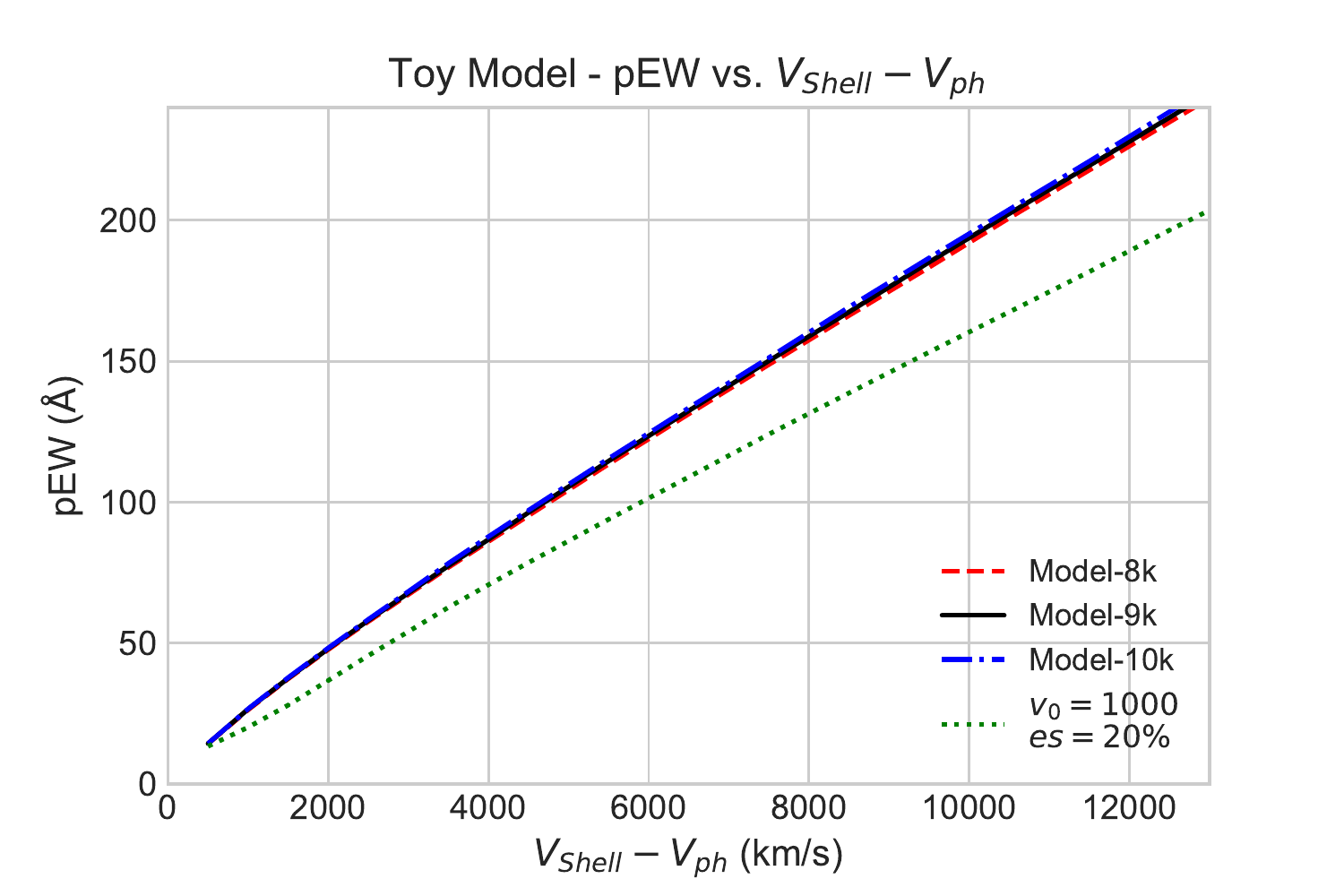}
	\caption{Relations between feature parameters in measured BSNIP \citep{Silverman2012} data compared with simple toy models with $V_{\rm ph}=8,9,10\times10^3$~km/s and a varying (step-function) $V_{\rm shell}$, and one toy model with $V_{\rm ph}=9\times10^3$~km/s and a smooth exponentially declining optical depth\added{ plus electron scattering flux} (see \S\ref{sec:extent}). \textbf{Top:} Fractional depth and FWHM vs. pEW \deleted{-}\added{--} BSNIP measurements for $6100\angstrom$ and $5750\angstrom$ features (blue and red dots) $\pm5$~days from peak and toy model results\deleted{ (solid and dotted lines)}. \textbf{Bottom left:} Resulting toy model spectra for $V_{\rm ph}=9\times10^3$~km/s with increasing $V_{\rm shell}$: First the fractional depth grows until it nears a maximum value. During this phase, the FWHM decreases. After the fractional depth is saturated, the FWHM begins to increase.\added{ The dotted line shows an example of the exponentially declining model for $V_{\rm shell}=23,500$~km/s.} \textbf{Bottom right:} In all toy models, the pEW of a feature grows almost linearly with the extent of the absorbing region $V_{\rm shell}-V_{\rm ph}$.\added{ The exponentially declining model has systematically lower pEWs due to the addition of 20\% white noise emulating electron scattering flux.}} 
	\label{fig:feature_formation}
	
\end{figure*}

The shape of an absorption line can be roughly described by two parameters\deleted{ -}\added{:} the maximal fractional depth (a) and the full-width at half-maximum (FWHM). The observed relations between these parameters and the pEW of the \ion{Si}{II} features are shown in the top panels of Fig.~\ref{fig:feature_formation} for the BSNIP \citep{Silverman2012} sample of \deleted{t}\added{T}ype Ia supernovae near peak ($\pm5$ days from peak). As can be seen, the FWHM and fractional depth are correlated with the pEW in a non-trivial way: for small pEWs ($\rm pEW\lesssim 100 \angstrom$), the FWHM is approximately constant while the depth grows linearly with the pEW. For large pEWs the depth saturates and the FWHM grows with the pEW. 

%%%%%%%%%%%%%%%%%%%%%%%%%%%%%%%%%%%%%%%%%%%%%%%%%%

%%%%%%%%%%%%%%%%%%%%%%%%%%%%%%%%%%%%%%%%%%%%%%%%%% 

In order to study these relations we consider a simple spherical toy model:
A photosphere expanding at $V_{\rm ph}=9,000$~km/s emits a flat spectrum and is surrounded by a shell with an infinite Sobolev optical depth $\tau_s=\infty$, with each interaction resulting in an (isotropic) scattering. Note that since photons are continuously red-shifted with respect to the rest-frame of the expanding plasma, each photon will (effectively) interact only once with a given transition. The shell extends from the photosphere to an outer velocity $V_{\rm shell}$ which sets the pEW of the line and is varied from $9,500$ \deleted{-}\added{to} $23,500$~km/s to account for the observed range of pEWs. The resulting absorption feature shapes are generated using a Monte-Carlo calculation ($5\times10^{8}$ photons per spectrum) and shown in the bottom-left panel of Fig.~\ref{fig:feature_formation}.

Here and throughout this study, feature parameters are extracted as follows (similar to the procedure described in \citealt{Silverman2012}): First, the spectrum is passed through a Savitzky-Golay filter \citep{SavGol64}. Next, the endpoints of the feature are located \deleted{-}\added{--} the feature is scanned from its minimum to both sides until a maximum is reached (with additional filtering at this stage). The pseudo-continuum is defined as a line passing through the two endpoints, and the spectrum is normalized by this pseudo-continuum (examples of pseudo-continuum lines can be seen in Fig. \ref{fig:Examples}). The pEW is determined by integrating the outcome of the normalized feature spectrum subtracted from unity. The fractional depth and FWHM are also extracted from the normalized spectrum.  

The resulting feature parameters obtained with this toy model are shown in the top panels of Fig.~\ref{fig:feature_formation} \deleted{in solid lines }for 3 choices of photospheric velocity $V_{\rm ph}$. \added{We suspect that the remaining difference between this simple model and the observations can be bridged by using a smoothly declining density profile, and introducing extra flux due to electron scattering in the ejecta. This is demonstrated by the dotted curve, derived from }\deleted{A dotted line shows }a similar model with $V_{\rm ph}=9,000$~km/s and a smooth exponentially declining optical depth of the form \deleted{$\tau = e^{-\frac{v-v_{\rm shell}}{v_0}}$}\added{$\tau = \frac{1}{2}e^{-\frac{v-v_{\rm shell}}{v_0}}$ with  $v_0=1000$~km/s}\deleted{ instead of a step function}\added{, along with an addition of 20\% white noise emulating flux due to electron scattering}. As can be seen, the toy model recovers the relations between the different absorption line parameters to a reasonable accuracy for \added{these choices}\deleted{the choice $v_0=1500$~km/s}. 

\subsection{Line pEW measures extent of absorption region}
\label{sec:extent2}

The fact that this simple model captures the main statistical features of the shapes of the absorption lines provides support for the approximation of a photosphere and a single absorbing line. Results of other models simulated with \textsc{Tardis} are shown in Fig.~\ref{fig:simulated}. All simulated models are consistent with the observed behavior.

As can be seen in the bottom-right panel of Fig.~\ref{fig:feature_formation}, in this simple model the pEW is insensitive to the entire distribution of \ion{Si}{II} ions and is a good estimator for the extent of the absorption region for the entire range of values (unlike the FWHM and fractional depths which saturate at the extreme regimes). 

Thus, to a first approximation, the pEW is a measure of the extent of the absorption region, within which the optical depth is larger than unity. In the following sections, we shall see how the extent of the absorption region of the \ion{Si}{II} features depends on the Si density profile and the luminosity of the supernova.

%%%%%%%%%%%%%%%%%%%%%%%%%%%%%%%%%%%%%%%%%%%%%%%%%%
%%%%%%%%%%%%%%%%%%%%%%%%%%%%%%%%%%%%%%%%%%%%%%%%%%
\section{Radiative transfer using Tardis}
\label{sec:radiative_transfer}

In order to explore the parameters affecting the Branch plot, we use the photospheric spectral synthesis code \textsc{Tardis} \citep{Tardis}. \textsc{Tardis} is a Monte-Carlo radiation transfer code that solves for a steady-state radiation field in a spherically symmetric ejecta with a predefined photosphere. Energy packets are injected at the photosphere with a black body distribution, propagate through the ejecta, interact with the plasma via bound-bound transitions and electron scattering and form the observed spectra when they escape. In our study, we use \textsc{Tardis} with the most detailed \texttt{macroatom} model. The ionization fractions and level populations are iteratively calculated using rough approximations for deviations from LTE as explained below. 

%%%%%%%%%%%%%%%%%%%%%%%%%%%%%%%%%%%%%%%%%%%%%%%%%%

\subsection{Non-LTE}
\label{sec:NLTE}
The ionization and excitation state of the plasma in \textsc{Tardis} is calculated (with some corrections) by assuming that the radiation field is given by a diluted black body:
%%%%%%%%%%%%%%%%%%%%%%%%%%%%%%%%%%%%%%%%%%%%%%%%%%
\begin{equation}
J_{\nu}=WB_{\nu}(T_R)
\label{eq:diluted}
\end{equation} 
%%%%%%%%%%%%%%%%%%%%%%%%%%%%%%%%%%%%%%%%%%%%%%%%%%
where the dilution factor $W$ and radiation temperature $T_R$ are calculated based on appropriate estimators from photon packets. The free electrons are assumed to have a temperature $T_e=0.9T_R$. The ionization is calculated by solving a modified Saha equation following the \texttt{nebular} ionization approximation (based on \citealt{ML93}) :
%%%%%%%%%%%%%%%%%%%%%%%%%%%%%%%%%%%%%%%%%%%%%%%%%%
\begin{equation}
    \frac{N_{i,j+1}n_e}{N_{i,j}}= D\left(\frac{N_{i,j+1}n_e}{N_{i,j}}\right)^{LTE(Saha)}
	\label{eq:Nebular}
\end{equation}
%%%%%%%%%%%%%%%%%%%%%%%%%%%%%%%%%%%%%%%%%%%%%%%%%%
where $D$ is an ansatz correction factor that depends on $W$, the electron to radiation temperature ratio $T_e/T_R$, the fraction of recombinations that go directly to the ground state for each ion and corrections to account for the dominance of locally created radiation at short wavelengths (see eq. 2 \deleted{-}\added{and} 3 in \citet{Tardis}).

The excitation levels are found using the \texttt{dilute-lte} excitation mode where the population of excited states is equal to the Boltzmann (LTE) population multiplied by the dilution factor $W$ (excluding metastable states which are set to the Boltzmann distribution). As an exception, the excitation levels of \ion{Ca}{II}, \ion{S}{II}, \ion{Mg}{II} and \ion{Si}{II} ions are calculated with a "full NLTE" treatment, by explicitly finding the steady-state solution to radiative and collisional excitation transition equations (using the diluted black-body radiation field and including correction factors for multiple interactions before escape for finite Sobolev optical depths). As shown in \citet{Tardis}, this has a significant effect on the pEW of the $5750\angstrom$ feature.

The effect of this approximate NLTE treatment on the level populations is calculated and presented in Fig.~\ref{fig:Saha} based on the \textsc{Tardis} implementation and atomic data by \citet{Kurucz} adapted from CMFGEN \citep{CMFGEN}, using a typical near-photospheric dilution factor of $W=0.3$. The resulting NLTE densities for the relevant levels are shown in solid lines. As can be seen in the figure, the NLTE treatment does not change the qualitative dependence of the level populations on temperature but has a significant quantitative effect. Specifically, the saturation effect discussed in \S\ref{sec:properties2} now presents at $\sim8000\rm\,K$.

%%%%%%%%%%%%%%%%%%%%%%%%%%%%%%%%%%%%%%%%%%%%%%%%%%
%%%%%%%%%%%%%%%%%%%%%%%%%%%%%%%%%%%%%%%%%%%%%%%%%%

\subsection{Ejecta from SNe Type Ia explosion models}

\subsubsection{1-D Chandrasekhar-mass and sub-Chandrasekhar models}
\label{sec:models}
We use \textsc{Tardis} to simulate radiative transfer through ejecta of spherically symmetric models from the literature. These include central detonations of sub-Chandrasekhar mass WD (SCH, \citealt{Blondin2017}) and delayed-detonations of Chandrasekhar-mass WD (DDC, \citealt{Blondin2013}) models with a range of $^{56}$Ni spanning the \deleted{t}\added{T}ype Ia range. Relevant model parameters are given in Table~\ref{table:models}. For simplicity, we use $V_{\rm ph}=9,000$~km/s as the photosphere velocity for all models (the exact value has limited effect on the qualitative analysis). Using parameters given by the above authors, we set the time from explosion to $t_{\rm rise}($bol$)$ and set $L_{\rm bol}^{\max}$ as the target luminosity for each model.

We extract the \ion{Si}{II} features' pEWs from both the original spectra (derived from the radiation transfer simulations presented in the original papers) and the spectra obtained using \textsc{Tardis}. The results of both methods are shown in the left panel of Fig.~\ref{fig:Branch2}. As can be seen in the figure, the results are qualitatively similar. While quantitative differences clearly exist, both methods place these models on the right side of the Branch plot, covering it only partially.

%%%%%%%%%%%%%%%%%%%%%%%%%%%%%%%%%%%%%%%%%%%%%%%%%%
\begin{table}
	\centering
	\begin{tabular}{ |c|c|c|c|c| } 
		\hline
		Model & $M_{\rm tot}$ & $M(^{56}$Ni$)$ & $t_{\rm rise}($bol$)$ & $L_{\rm bol}^{\max}$ \\
		& ($\textup{M}_\odot$) & ($\textup{M}_\odot$) & (days) & (erg/s) \\
		\hline
		\multicolumn{5}{|c|}{Chandrasekhar-mass delayed-detonation models} \\
		\hline
		DDC0  & 1.41 & 0.86 & 16.7 & 1.85 (43)\\ 
		DDC6  & 1.41 & 0.72 & 16.8 & 1.57 (43)\\ 
		DDC10 & 1.41 & 0.62 & 17.1 & 1.38 (43)\\ 
		DDC15 & 1.41 & 0.51 & 17.6 & 1.14 (43)\\ 
		DDC17 & 1.41 & 0.41 & 18.6 & 9.10 (42)\\ 
		DDC20 & 1.41 & 0.30 & 18.7 & 6.65 (42)\\ 
		DDC22 & 1.41 & 0.21 & 19.6 & 4.47 (42)\\ 
		DDC25 & 1.41 & 0.12 & 21.0 & 2.62 (42)\\ 
		\hline
		\multicolumn{5}{|c|}{Sub-Chandrasekhar-mass models}\\
		\hline
		SCH7p0 & 1.15 & 0.84 & 16.4 & 1.85 (43)\\
		SCH6p5 & 1.13 & 0.77 & 16.6 & 1.71 (43)\\
		SCH6p0 & 1.10 & 0.70 & 16.9 & 1.57 (43)\\
		SCH5p5 & 1.08 & 0.63 & 17.1 & 1.42 (43)\\
		SCH5p0 & 1.05 & 0.55 & 17.6 & 1.25 (43)\\
		SCH4p5 & 1.03 & 0.46 & 17.7 & 1.08 (43)\\
		SCH4p0 & 1.00 & 0.38 & 17.6 & 9.01 (42)\\
		SCH3p5 & 0.98 & 0.30 & 17.2 & 7.34 (42)\\
		SCH3p0 & 0.95 & 0.23 & 16.8 & 5.76 (42)\\
		SCH2p5 & 0.93 & 0.17 & 16.5 & 4.36 (42)\\
		SCH2p0 & 0.90 & 0.12 & 15.8 & 3.17 (42)\\
		SCH1p5 & 0.88 & 0.08 & 15.0 & 2.26 (42)\\
		\hline
	\end{tabular}
	\caption{Properties of SNe Ia models adapted from \citet{Blondin2017}. Numbers in parentheses correspond to powers of ten.}
	\label{table:models}
\end{table}
%%%%%%%%%%%%%%%%%%%%%%%%%%%%%%%%%%%%%%%%%%%%%%%%%%

%%%%%%%%%%%%%%%%%%%%%%%%%%%%%%%%%%%%%%%%%%%%%%%%%%
\subsubsection{2-D Head-on collision models}
\label{sec:collision}

We use \textsc{Tardis} to simulate radiative transfer through ejecta derived from 2-D hydrodynamic simulations of head-on (zero impact parameter) collisions of CO-WDs. 
In \citet{Kushnir2013}, collisions of (equal and non-equal mass) CO-WDs with masses 0.5, 0.6, 0.7, 0.8, 0.9 and 1.0 $\textup{M}_\odot$ were simulated, resulting in explosions that synthesize $^{56}$Ni masses in the range of $0.1\,\textup{M}_\odot$ to $1.0\,\textup{M}_\odot$. The models used in this study (excluding rare collisions with $1.0\,\textup{M}_\odot$ WDs) are summarized in Table~\ref{table:collisions}.

\begin{figure}
	\centering
	\includegraphics[clip, trim=4.5cm 0cm 5.3cm 0cm, width=0.38\columnwidth]{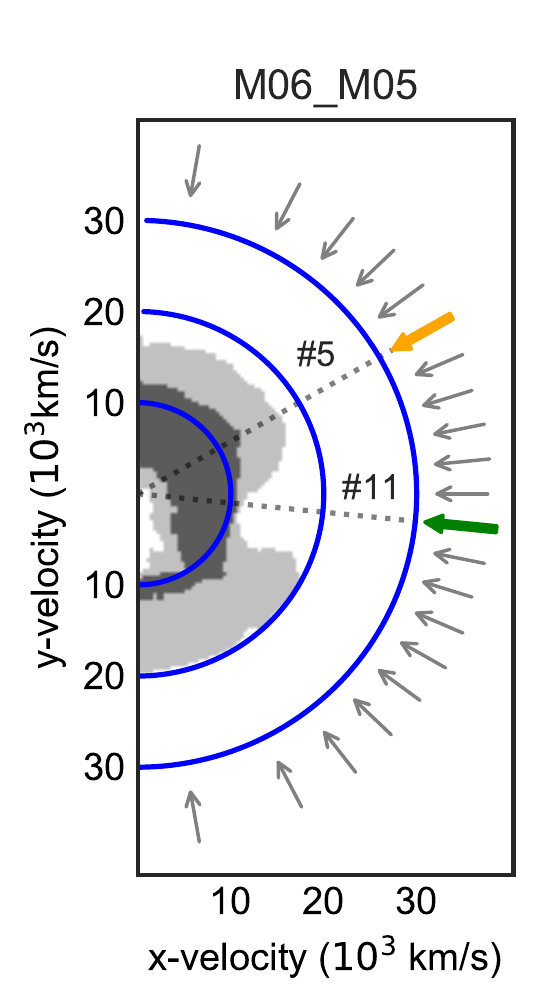}
	\caption{Si density map of a head-on collision model ejecta. Here a $0.6\,\textup{M}_\odot$ WD collides with a $0.5\,\textup{M}_\odot$ WD. The collision axis is the y-axis. The arrows represent simulated viewing directions. The orange and green arrows represent viewing angles 5 and 11 in correspondence with Fig.~\ref{fig:Branch2} and Fig.~\ref{fig:Density_Models}. The darker area represents regions with Si density > $8\times 10^{-15}$\gcm at $18.0$ days after explosion. The lighter area is the same with Si density > $6\times 10^{-16}$\gcm (see \S\ref{sec:threshold}).}
	\centering
	\label{fig:slices}
\end{figure}

In order to use the 1-D \textsc{Tardis} package, each collision model ejecta was sliced into 21 viewing angles. An example can be seen in Fig.~\ref{fig:slices}. In this example it is apparent that the collision model produces an asymmetric ejecta: for some viewing angles, the Si extends to $20,000$~km/s, while for others the Si density drops steeply at approximately $12,000$~km/s.
From each viewing angle, a 1-D model was generated with the density and abundance values sampled along the section. The luminosity $L_{\rm bol}^{\max}$ and rise time $t_{\rm rise}($bol$)$ were taken for each viewing angle from global 2-D LTE radiation transfer simulations of the same ejecta performed by Wygoda, N. (private communication) using a 2-D version of the URILIGHT radiation transfer code \citep[][Appendix A]{Wygoda.2}. Average values are given in Table~\ref{table:collisions}, and detailed values are available in the supplementary files. The inner boundary of the simulation was set at $V_{\rm ph}=9,000$~km/s for all models. 

The resulting Branch plot distribution is shown in the right panel of Fig.~\ref{fig:Branch2}. Remarkably, we find that the model covers most of the observed Branch plot distribution. A comparison of two models with the same $^{56}$Ni yield but different Si density profiles (\texttt{M06\_M05\_11} and \texttt{M06\_M05\_5}, see Figs.~\ref{fig:Density_Models} and \ref{fig:slices}) shows the correspondence between the Si density profile and the position on the Branch plot. 
Examples of resulting spectra compared to observed spectra are presented in Appendix \ref{sec:Spectra}.

We note that the construction presented here is not assumed to be exact. The method of taking a section of a 2-D model and producing from it a 1-D model is not equivalent to 2-D radiation transfer. However, the shape of absorption features depends mainly on the composition of material in the line of sight. Thus, the qualitative result that the collision model covers most of the Branch plot will likely hold. Additionally, actual collisions will cover a range of impact parameters and result in 3-D ejecta that are unavailable at this time.

%%%%%%%%%%%%%%%%%%%%%%%%%%%%%%%%%%%%%%%%%%%%%%%%%%

\begin{table}
    \centering
    \begin{tabular}{ |c|c|c|c|c|c| } 
    \hline
    Model & $M_1$              & $M_2$             & $M(^{56}$Ni$)$    & $t_{\rm rise}($bol$)$ & $L_{\rm bol}^{\max}$ \\
          & ($\textup{M}_\odot$)  & ($\textup{M}_\odot$) & ($\textup{M}_\odot$)  & (days) & (erg/s) \\
    \hline
    \multicolumn{6}{|c|}{Head-on collision models} \\
    \hline
    M05\_M05 & 0.50 & 0.50 & 0.10 & 14.8 & 2.98 (42)\\
    M055\_M055 & 0.55 & 0.55 & 0.22 & 15.7 & 5.56 (42)\\
    M06\_M05 & 0.60 & 0.50 & 0.27 & 15.3 & 6.54 (42)\\
    M06\_M06 & 0.60 & 0.60 & 0.33 & 16.0 & 7.64 (42)\\
    M07\_M05 & 0.70 & 0.50 & 0.26 & 15.7 & 6.51 (42)\\
    M07\_M06 & 0.70 & 0.60 & 0.38 & 16.0 & 8.81 (42)\\
    M07\_M07 & 0.70 & 0.70 & 0.56 & 15.9 & 1.25 (43)\\
    M08\_M05 & 0.80 & 0.50 & 0.29 & 16.2 & 7.32 (42)\\
    M08\_M06 & 0.80 & 0.60 & 0.38 & 16.3 & 9.54 (42)\\
    M08\_M07 & 0.80 & 0.70 & 0.48 & 16.5 & 1.17 (43)\\
    M08\_M08 & 0.80 & 0.80 & 0.74 & 15.5 & 1.67 (43)\\
    M09\_M05 & 0.90 & 0.50 & 0.69 & 15.6 & 1.34 (43)\\
    M09\_M06 & 0.90 & 0.60 & 0.50 & 16.5 & 1.26 (43)\\
    M09\_M07 & 0.90 & 0.70 & 0.51 & 16.7 & 1.23 (43)\\
    M09\_M08 & 0.90 & 0.80 & 0.54 & 17.1 & 1.27 (43)\\
    M09\_M09 & 0.90 & 0.90 & 0.78 & 16.8 & 1.74 (43)\\
    \hline
    \end{tabular}
    \caption{Properties of SNe Ia head-on collision models adapted from \citet{Kushnir2013}. Numbers in parentheses correspond to powers of ten. $L_{\rm bol}^{\max}$ and $t_{\rm rise}($bol$)$ represent averages over all viewing angles (see \S\ref{sec:collision}). Detailed values are provided in the supplementary files and show a standard deviation of $\sim5\%$ for $t_{\rm rise}($bol$)$ and $\sim10\%$ for $L_{\rm bol}^{\max}$ across viewing angles.}
    \label{table:collisions}
\end{table}

%%%%%%%%%%%%%%%%%%%%%%%%%%%%%%%%%%%%%%%%%%%%%%%%%%
%%%%%%%%%%%%%%%%%%%%%%%%%%%%%%%%%%%%%%%%%%%%%%%%%%

%%%%%%%%%%%%%%%%%%%%%%%%%%%%%%%%%%%%%%%%%%%%%%%%%%

%%%%%%%%%%%%%%%%%%%%%%%%%%%%%%%%%%%%%%%%%%%%%%%%%%

\subsection{Exploring synthetic ejecta}
\label{sec:synthetic}

\begin{figure*}
	\centering
	\includegraphics[width=\columnwidth]{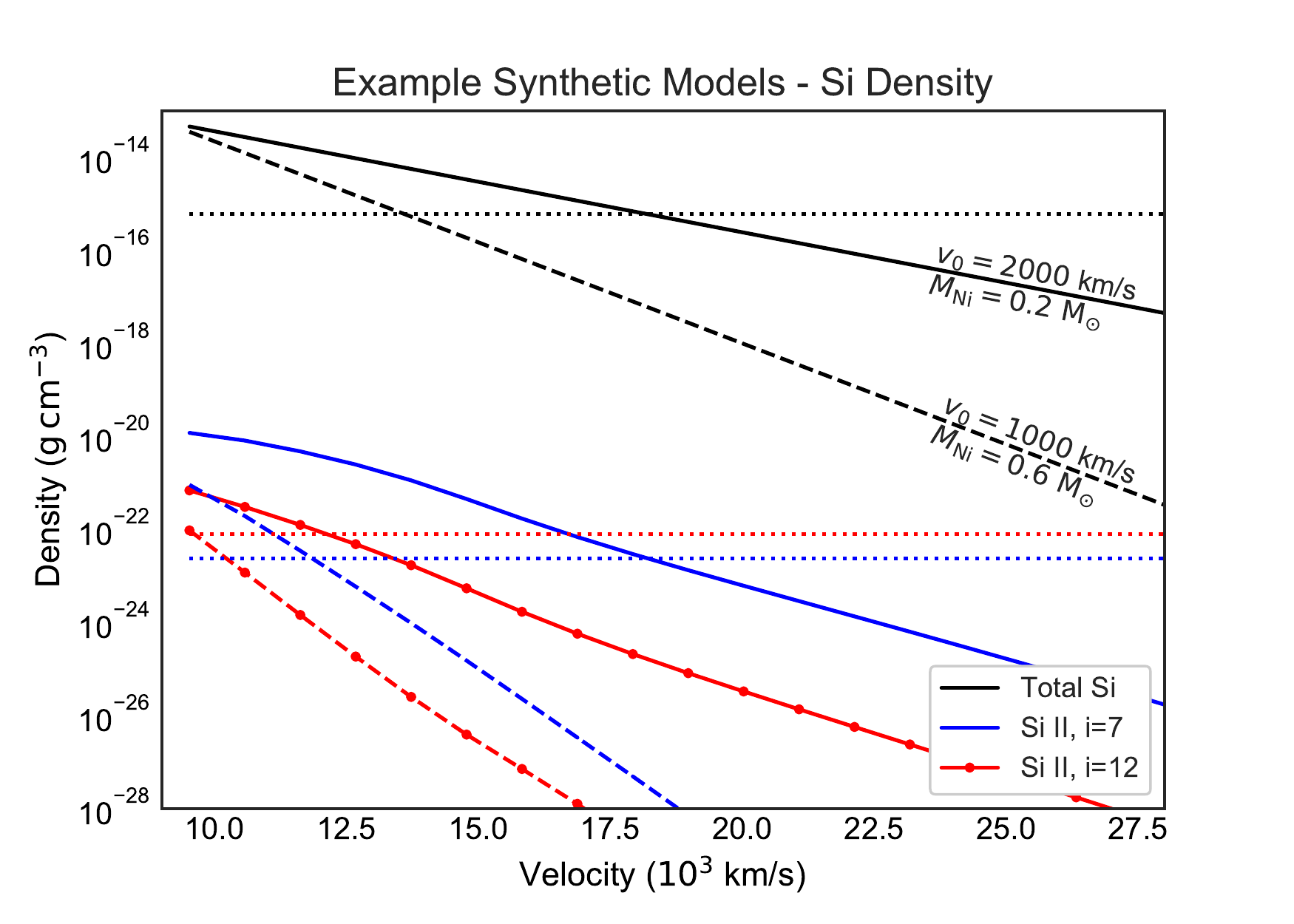}
	~
	\includegraphics[width=\columnwidth]{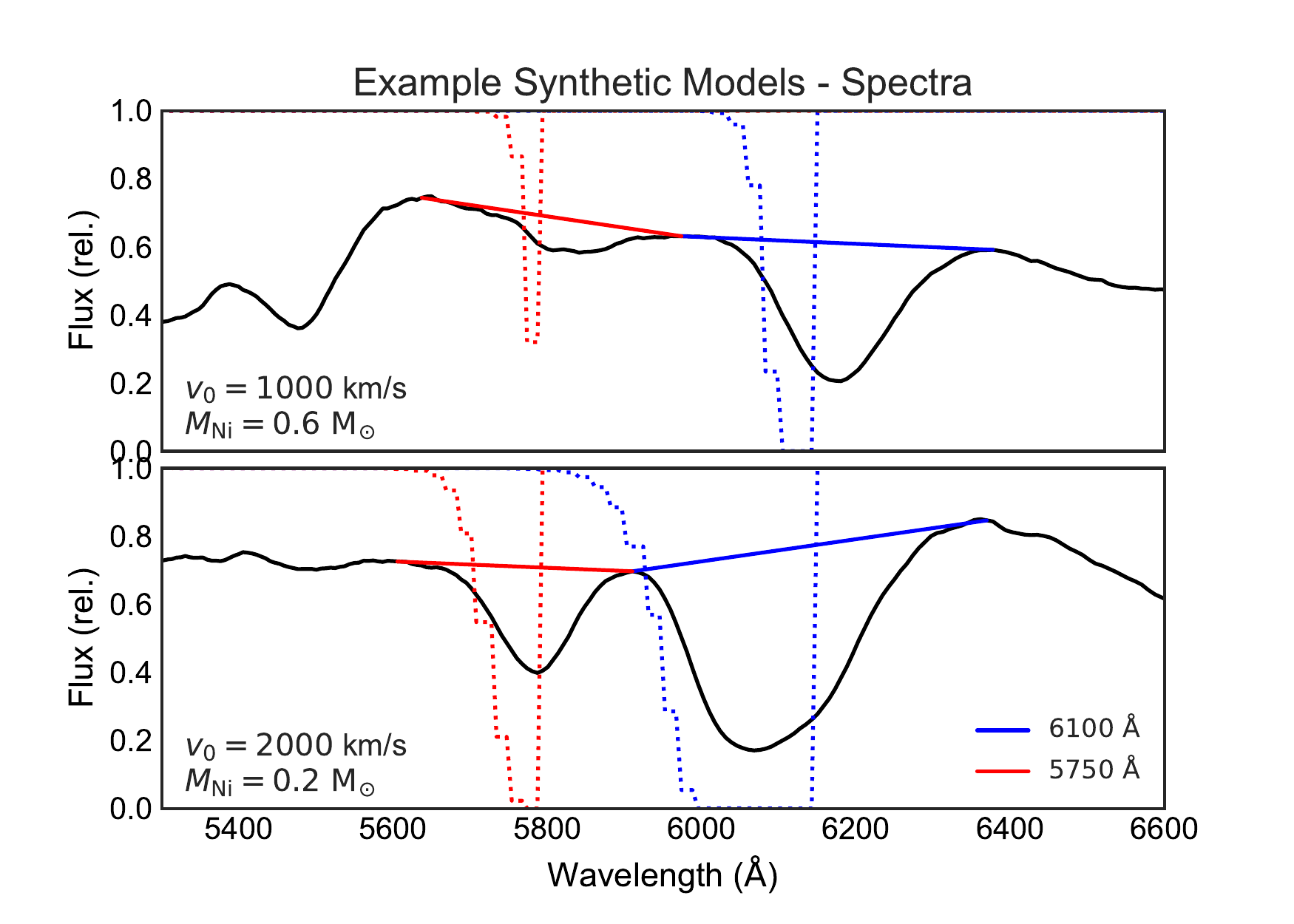}
	\caption{Results of \textsc{Tardis} radiative transfer simulations on synthetic exponential density models of the form $\rho ({v})=\rho_0 e^{-v/v_0}$, $\rho=10^{-13}$\gcm~at $V_{\rm ph}=9,000$~km/s at 18 days. Two models are shown as examples (marked with black circles in the left panel of Fig.~\ref{fig:Branch}). \deleted{- \textbf{Left:} Low e-folding constant $v_0=1,000$~km/s with a high $M(^{56}$Ni$)=0.6\,\textup{M}_\odot$. \textbf{Right:} High $v_0=2,000$~km/s with a low $M(^{56}$Ni$)=0.2\,\textup{M}_\odot$. Panels show the attenuation profile, density and temperature within the ejecta. The spectrum is also shown, including the automatic detection of the pseudo-continuum for the pEW calculation, with the original attenuation profile that generated the spectral features overlaid. In the density panel, thresholds obtained in \S\ref{sec:threshold} are shown in dashed lines, and dotted blue and red lines represent the necessary $n_l$ for $\tau_s=1$ without stimulated emission (see Eqs.~\ref{eq:sobolev}). The lower temperature at $V_{\rm ph}$ in the example on the right, results in a higher population of both levels. Both this and the higher e-folding constant $v_0$ cause the absorption region to be more extended than in the example on the left, resulting in larger pEWs.}\added{\textbf{Left:} Total Si density and density of \mbox{\ion{Si}{II}} ions excited to the relevant levels. Solid lines represent a model with $v_0=2,000$~km/s, $M(^{56}$Ni$)=0.2\,\textup{M}_\odot$ and dashed lines represent a model with $v_0=1,000$~km/s, $M(^{56}$Ni$)=0.6\,\textup{M}_\odot$. Dotted blue and red lines represent the necessary $n_l$ for $\tau_s=1$ without stimulated emission (see Eqs.~\mbox{\ref{eq:sobolev}}). The dotted black line represents the total Si density threshold obtained in \S\mbox{\ref{sec:threshold}} for the $6100\angstrom$ feature. \textbf{Right:} Simulated spectra, including the automatic detection of the pseudo-continuum for the pEW calculation, with the original attenuation profile that generated the spectral features overlaid (dotted lines). The lower temperature at $V_{\rm ph}$ in the $v_0=2,000$~km/s, $M(^{56}$Ni$)=0.2\,\textup{M}_\odot$ model results in a higher population of both levels. Both this and the higher e-folding constant $v_0$ cause the absorption region to be more extended than in the second model, resulting in larger pEWs for both features.}} 
	\label{fig:Examples}
\end{figure*}

In this section, we apply the \textsc{Tardis} radiative transfer simulation to synthetic ejecta models. This allows an exploration of the dependency of the Branch plot distribution on the properties of the ejecta in a controlled manner. We use simple exponentially declining density models of the form: $\rho ({v})=\rho_0 e^{-{v/v_0}}$ up to a maximal velocity $v_{\max}$. We use the same elemental abundance as in \S\ref{sec:properties2}. We maintain a typical density for ejecta at maximum light at the photosphere ($\rho=10^{-13}$\gcm~at $V_{\rm ph}=9,000$~km/s at 18 days). We vary: (1) the target output luminosity (equivalent to $0.15<M(^{56}$Ni$)/\textup{M}_\odot<1.0$); (2) the e-folding of the exponent ($500$~km/s~$<{v_0}<2,500$~km/s) and (3) the maximum velocity of the ejecta ($10,000$~km/s~$<v_{\max}<30,000$~km/s). Fig.~\ref{fig:Examples} shows example results of two such models.

%%%%%%%%%%%%%%%%%%%%%%%%%%%%%%%%%%%%%%%%%%%%%%%%%

%%%%%%%%%%%%%%%%%%%%%%%%%%%%%%%%%%%%%%%%%%%%%%%%%

Fig.~\ref{fig:Branch} shows the resulting Branch plots overlaid on observed data. In the left panel, the maximum ejecta velocity is kept constant at $v_{\max}=30,000$~km/s. The different lines connect points of constant $v_0$ and varying luminosity (represented as $M(^{56}$Ni$)/\textup{M}_\odot$). In the right panel, the e-folding velocity is kept constant at $v_0=2,500$~km/s. The different lines connect points of constant $v_{\max}$ and varying luminosity. Additional models in which only the Si density profile varies while the total density profile remains constant show very similar results (see Fig.~\ref{fig:Branch_Si}). We identify several interesting effects: 

%%%%%%%%%%%%%%%%%%%%%%%%%%%%%%%%%%%%%%%%%%%%%%%%%%

\begin{figure*}
	\centering
	\includegraphics[width=\columnwidth]{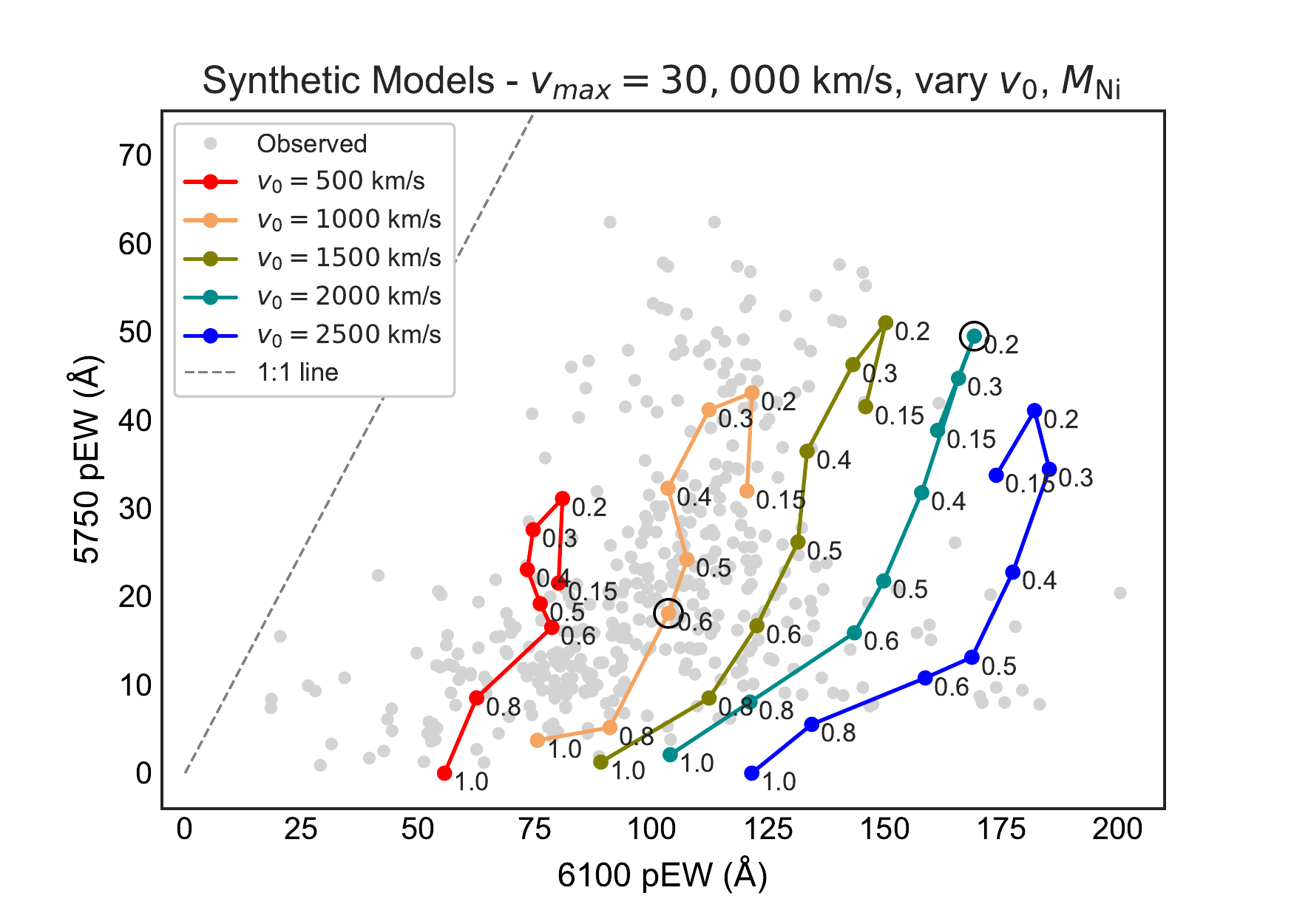}
	~ 
	\includegraphics[width=\columnwidth]{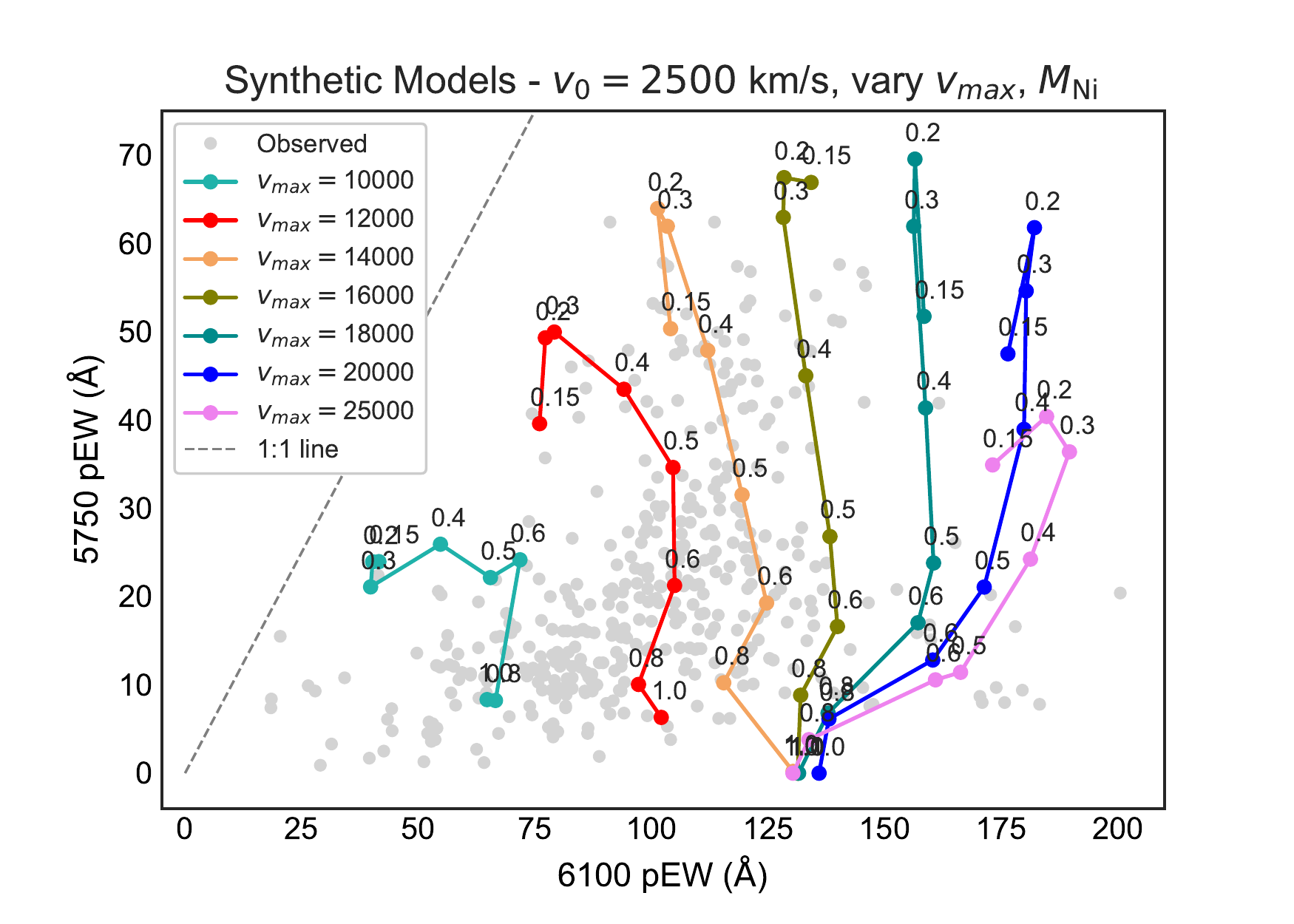}
	\caption{Simulated synthetic models with exponential density profiles of the form $\rho ({v})=\rho_0 e^{-v/v_0}$, $\rho=10^{-13}$\gcm~at $V_{\rm ph}=9,000$~km/s at 18 days, overlaid on observed CfA, CSP and BSNIP data. $M(^{56}$Ni$)$ is stated next to each point in units of $\textup{M}_\odot$. $L$ is computed from $M(^{56}$Ni$)$ using Arnett's rule $L_{\max}=2.0\times10^{43}\times[M(^{56}$Ni$)/\textup{M}_\odot]$~erg/s. \textbf{Left:} Each line represents a constant e-folding velocity $v_0$ with varying luminosity. All models are simulated to $v_{\max}=30,000$~km/s. Circles mark models corresponding to examples in Fig.~\ref{fig:Examples}. \textbf{Right:} Exponential synthetic models with a constant e-folding velocity $v_0=2,500$~km/s. Each line represents a constant maximum ejecta velocity cutoff $v_{\max}$ with varying luminosity.}
	
	\label{fig:Branch}
\end{figure*}

%%%%%%%%%%%%%%%%%%%%%%%%%%%%%%%%%%%%%%%%%%%%%%%%%%

\subsubsection{Luminosity explains one dimension of the plot}
\label{sec:lumin_explains}

Looking at both panels of Fig.~\ref{fig:Branch}, we see that for higher luminosities ($M(^{56}$Ni$)/\textup{M}_\odot>0.2$), decreasing the target luminosity increases the $5750\angstrom$ pEW and allows the model to climb higher in the Branch plot (see also \citealt{Heringer2017}). 

Fig.~\ref{fig:Examples} provides an instructive example of how the luminosity affects the extent of the absorption region and the pEW of the features. In the \added{$v_0=2,000$~km/s, $M(^{56}$Ni$)=0.2\,\textup{M}_\odot$ model}\deleted{example on the right}, the luminosity is low, and the temperature at the photosphere is $\sim8000\rm\,K$. The corresponding level populations are well above the $\tau_s=1$ threshold (see \S\ref{sec:properties2}), and thus the attenuation profiles of both features begin at 100\%. In contrast, in the \added{$v_0=1,000$~km/s, $M(^{56}$Ni$)=0.6\,\textup{M}_\odot$ model}\deleted{example on the left}, the luminosity is high, and the temperature at the photosphere is $\sim10,000\rm\,K$. As a result, the level populations are lower: the $6100\angstrom$ feature begins at 100\% attenuation, but the $5750\angstrom$ feature begins at only 70\% attenuation and drops quickly. 

%%%%%%%%%%%%%%%%%%%%%%%%%%%%%%%%%%%%%%%%%%%%%%%%%%

\subsubsection{Lowering the luminosity further saturates both features}
\label{sec:low_lumin}

What happens when we further decrease the luminosity? As we can see in Fig.~\ref{fig:Branch}, it turns out that both features' pEW is limited due to the saturation effect shown in Fig.~\ref{fig:Saha}. When the temperature near the photosphere declines to values lower than $\sim8000\rm\,K$, the ionization levels out but the excitation continues to decrease. For sufficiently low luminosities, the temperature at the photosphere drops below $\sim8000\rm\,K$, thus reducing the level populations, the optical depth, and ultimately the pEW of the features. A similar effect can also be identified in the DDC and SCH models shown in the left panel of Fig.~\ref{fig:Branch2}.

%%%%%%%%%%%%%%%%%%%%%%%%%%%%%%%%%%%%%%%%%%%%%%%%%%

\subsubsection{Varying the e-folding velocity spans the width of the plot}
\label{sec:vary_v0}

Looking at the left panel of Fig.~\ref{fig:Branch}, we see that as the e-folding velocity $v_0$ is increased, the $6100\angstrom$ pEW reaches higher values (\citealt{Branch2009} obtained similar results using \textsc{Synow}). This can be explained using Fig.~\ref{fig:Examples}\deleted{-}\added{:} in the \added{$v_0=2,000$~km/s, $M(^{56}$Ni$)=0.2\,\textup{M}_\odot$ model}\deleted{model on the right}, the e-folding velocity $v_0$ is higher than in the \added{$v_0=1,000$~km/s, $M(^{56}$Ni$)=0.6\,\textup{M}_\odot$ model}\deleted{model on the left}, thus the Si density profile is more extended. Consequently, the level populations go below the $\tau_s=1$ threshold at a higher velocity, resulting in larger pEWs.

Are all of these density profiles physical? For an exponential density profile model for homologously expanding supernovae, $E = 6M v^2_0$ \citep[e.g.][]{Jeffery1999}. Assigning a typical energy to mass ratio for nuclear processes $\frac{E}{M}\approx\frac{0.5~\rm MeV}{\rm m_p}$, we obtain $v_0\approx2,800$~km/s. According to Fig.~\ref{fig:Branch}, this means that spherically symmetric models with an exponential density profile and constant abundance will tend to the right side of the Branch plot and will not be able to span the whole observed distribution. If this exponential model is representative, in order to cover the left side of the Branch plot, a model needs to produce viewing angles in which the Si density drops off more steeply than the total density. 

%%%%%%%%%%%%%%%%%%%%%%%%%%%%%%%%%%%%%%%%%%%%%%%%%%

\subsubsection{Limiting the maximum velocity of the ejecta spans the width of the plot}
\label{sec:limit_vshell}

Another demonstration of the ability to span the horizontal dimension of the plot by narrowing the absorption region is achieved by taking a synthetic ejecta with $v_0=2,500$~km/s (close to the $v_0\approx2,800$~km/s obtained in the previous section) and cutting it off at various velocities $v_{\max}$. The results of this approach are shown in the right panel of Fig.~\ref{fig:Branch}. 

Noticing that this method achieves a fuller coverage of the observed Branch plot, we conclude that the top left side of the plot, namely a high $5750\angstrom$ pEW with a low $6100\angstrom$ pEW, can be reached with this type of exponential model if a cut off in the Si profile exists at $v_{\max}\approx12,000$~km/s. 

%%%%%%%%%%%%%%%%%%%%%%%%%%%%%%%%%%%%%%%%%%%%%%%%%%

\section{Si density thresholds}
\label{sec:threshold}

\begin{figure}
	\centering
	\includegraphics[width=\columnwidth]{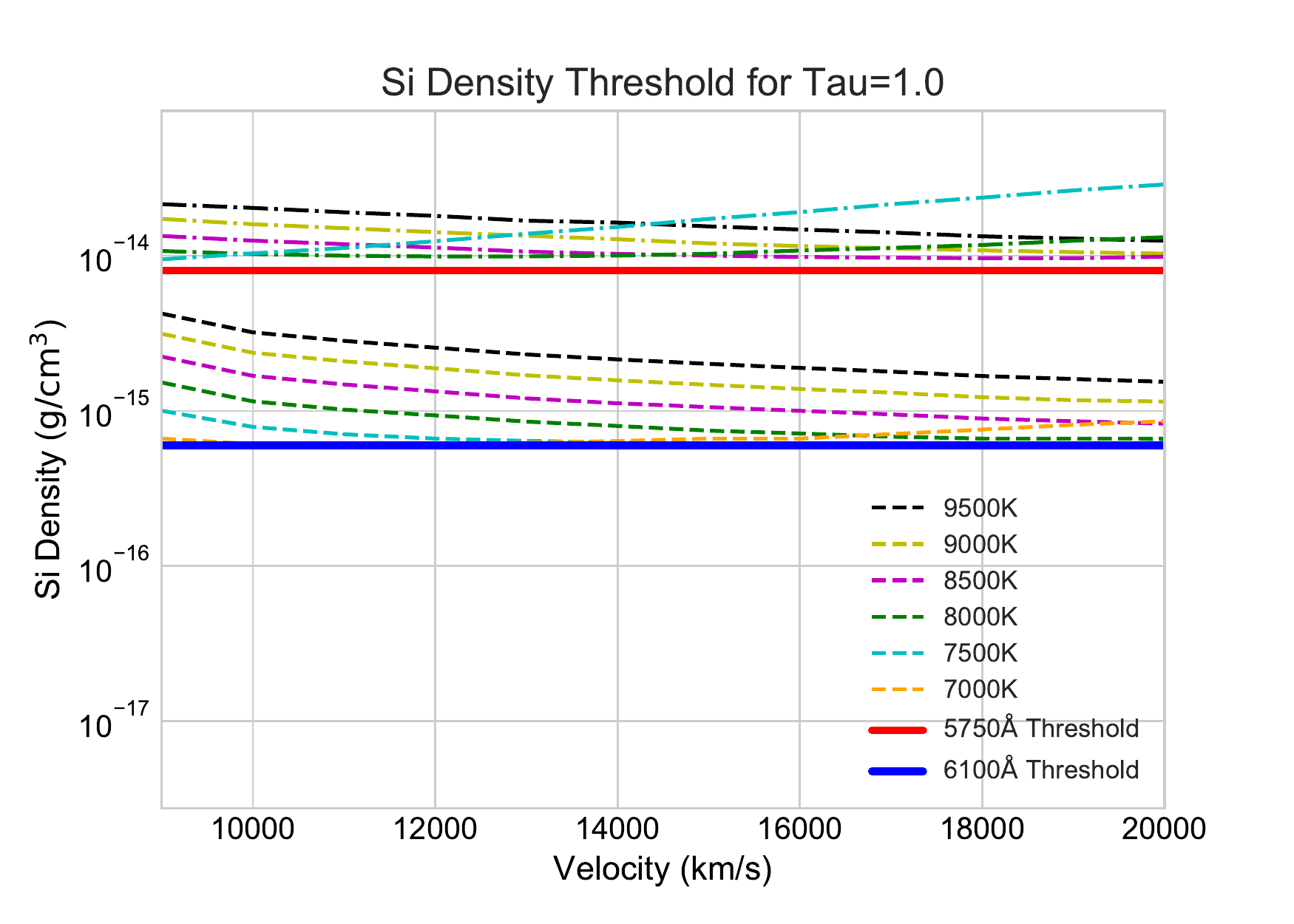}
	\caption{Total Si density necessary to obtain $\tau_s=1$ is iteratively computed for different temperatures and velocities, assuming a geometric dilution factor ($W=[1-(1-(V_{\rm ph}/{v})^2]^{1/2}/2$) within the ejecta. Temperatures shown for dashed lines for the $6100\angstrom$ feature, apply to same colored dash-dotted lines for the $5750\angstrom$ feature. Minimal thresholds ($\rho_{5750}$, $\rho_{6100}$) are marked in bold lines.}
	\label{fig:density}
\end{figure}

Using the NLTE model described in \S\ref{sec:NLTE}, we numerically find the Si density corresponding to $\tau_s=1$ for various temperatures and velocities in the ejecta, assuming a dilution factor with geometric dependence on velocity $W=[1-(1-(V_{\rm ph}/{v})^2]^{1/2}/2$. The results are shown in dashed ($6100\angstrom$) and dashed-dotted ($5750\angstrom$) lines in Fig.~\ref{fig:density}.

%%%%%%%%%%%%%%%%%%%%%%%%%%%%%%%%%%%%%%%%%%%%%%%%%%

%%%%%%%%%%%%%%%%%%%%%%%%%%%%%%%%%%%%%%%%%%%%%%%%%%

For low luminosity models, within the range $7,000\rm\,K$ \deleted{-}\added{to} $9,500\rm\,K$ and velocities $10,000~ {\rm km/s}<{v}< 20,000~ {\rm km/s}$, the Si density required for $\tau_s=1$ spans only one order of magnitude. Thus, we attempt to set total Si density thresholds of 
%%%%%%%%%%%%%%%%%%%%%%%%%%%%%%%%%%%%%%%%%%%%%%%%%%
%\begin{equation}
$\rho_{6100}=6\times 10^{-16}\rm g~cm^{-3}$
%\label{eq:threshold6100}
%\end{equation}
%%%%%%%%%%%%%%%%%%%%%%%%%%%%%%%%%%%%%%%%%%%%%%%%%%
for the $6100\angstrom$ feature, and 
%%%%%%%%%%%%%%%%%%%%%%%%%%%%%%%%%%%%%%%%%%%%%%%%%%
%\begin{equation}
$\rho_{5750}=8\times 10^{-15}\rm g~cm^{-3}$
%\label{eq:threshold5750}
%\end{equation}
%%%%%%%%%%%%%%%%%%%%%%%%%%%%%%%%%%%%%%%%%%%%%%%%%%
for the $5750\angstrom$ feature (solid lines in Fig.~\ref{fig:density}) for obtaining significant optical depth, in the hope that they are applicable to most relevant conditions. 

\begin{figure}
	\centering
	\includegraphics[clip, trim=0cm 0cm 2.5cm 0cm, width=\columnwidth]{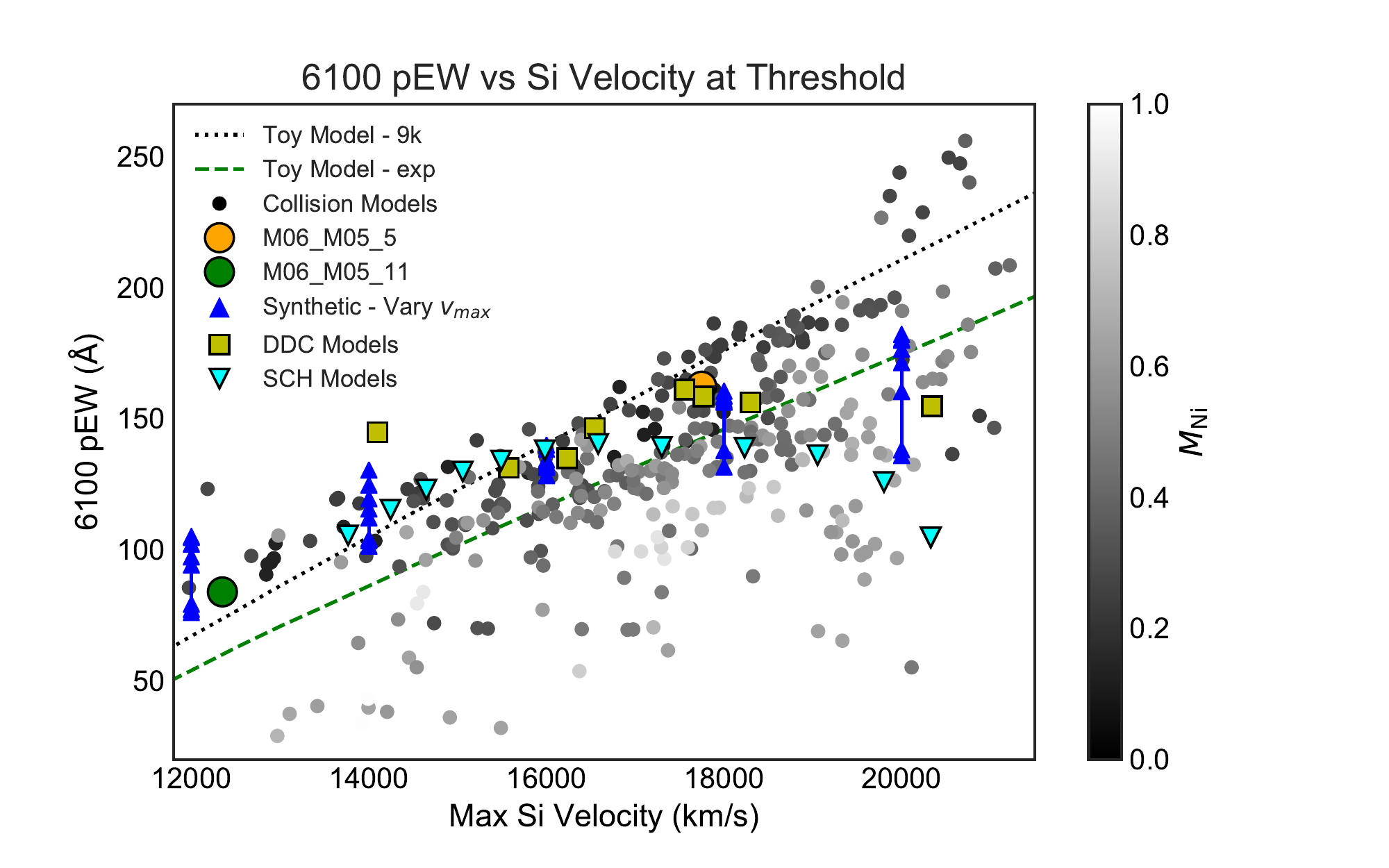}
	\caption{$6100\angstrom$ pEW vs effective "maximum Si velocity" (at which Si density = $\rho_{6100}$, see \S\ref{sec:threshold}) for various models. The head-on collision models are colored according to their $M(^{56}$Ni$)$. Orange and green circles show the two extreme models in correspondence with Figures~\ref{fig:Branch2}, \ref{fig:Density_Models} and \ref{fig:slices}. The dotted\deleted{ blue line is the relation between feature pEW and shell velocity from the $V_{\rm ph}=9,000$~km/s}\added{ and dashed lines represent the $V_{\rm ph}=9,000$~km/s and exponential} toy model\added{s} in \S\ref{sec:toy_model}.}
	
	\label{fig:Velocity}
\end{figure}

In order to test the predictive power of the thresholds obtained above, we find the intersection of the $\rho_{6100}$ threshold with the Si density profiles of all of the models presented in this paper to obtain effective "maximum Si velocities". These are plotted against the $6100\angstrom$ pEWs in Fig.~\ref{fig:Velocity}. The dotted line in the figure represents the $V_{\rm ph}=9,000$~km/s toy model of \S\ref{sec:toy_model}. The head-on collision and synthetic models show a correlation similar to the toy model, especially for low luminosity models, as would be expected from Fig.~\ref{fig:density}. On the other hand, the delayed-detonation and sub-Chandrasekhar models do not display a similar correlation. In both of these models, the luminosity is correlated with the extent of the Si density profile, and so they lack ejecta with high "maximum Si velocity" and low luminosity. A possible solution is defining a threshold that is a function of luminosity.

Also shown in Fig.~\ref{fig:Velocity} are green and orange circles representing two viewing angles of the \texttt{M06\_M05} head-on collision model (see Fig.~\ref{fig:slices}). Fig.~\ref{fig:Density_Models} shows the Si density profiles of these models and their intersection with the $\rho_{6100}$ threshold. \texttt{M06\_M05\_11} (in green) drops steeply and intersects $\rho_{6100}$ at $\sim12,000$~km/s whereas \texttt{M06\_M05\_5} (in orange) is more extended. A comparison with the right panel of Fig.~\ref{fig:Branch2} shows the correspondence between the intersection velocity and the position on the Branch plot. 

%%%%%%%%%%%%%%%%%%%%%%%%%%%%%%%%%%%%%%%%%%%%%%%%%%

%%%%%%%%%%%%%%%%%%%%%%%%%%%%%%%%%%%%%%%%%%%%%%%%%%

A similar exercise for the $\rho_{5750}$ threshold does not produce useful results. The Si density profiles tend to be flat close to the photosphere where this feature is formed (see Fig.~\ref{fig:Density_Models}) and the uncertainty in the threshold entails a large variation in the effective "maximum Si velocity".

%%%%%%%%%%%%%%%%%%%%%%%%%%%%%%%%%%%%%%%%%%%%%%%%%%

\section{Boundaries of the Branch plot}
\label{sec:boundaries}

Based on the results of the previous sections, we can now attempt to explain the shape and boundaries of the observed Branch plot.

%%%%%%%%%%%%%%%%%%%%%%%%%%%%%%%%%%%%%%%%%%%%%%%%%%

\subsection{Top boundary}
\label{sec:top_boundary}
It is well known \citep[e.g.][\S \ref{sec:lumin_explains}]{Nugent1995,Heringer2017} that as the SN luminosity decreases, the pEW of the $5750\angstrom$ feature increases. It was shown however in \S\ref{sec:low_lumin} that there is a limit on the maximum pEW of the $5750\angstrom$ feature due to the peak population of the \ion{Si}{II} $i=11,12$ levels at $T\sim8000\rm\,K$ (see Fig.~\ref{fig:Saha}). Looking at Fig.~\ref{fig:Branch}, it seems that the largest $5750\angstrom$ pEWs obtained with synthetic models generally match the largest observed pEWs, hinting that the maximal observed value may be related to this limit. On the other hand, ejecta with $^{56}$Ni masses of $\sim 0.1\,\textup{M}_{\odot}$ at the lower end of the \deleted{t}\added{T}ype Ia brightness distribution have comparable $5750\angstrom$ pEWs. 

Thus, it is hard to distinguish between the top boundary of the observed Branch plot distribution as being (a) due to the physical limit on the lowest luminosity of SNe Ia events; or (b) due to the optical depth reaching a maximum value at some temperature as a result of the atomic physics.

Currently, plots of the \ion{Si}{II} features' pEW vs. luminosity tracers such as $\Delta \rm m_{15}(\rm B)$ do not show a significant saturation effect (see e.g. Fig. 17 in \citealt{CSP2013}). Observations of lower luminosity SNe Ia may discover such an effect in the future.

%%%%%%%%%%%%%%%%%%%%%%%%%%%%%%%%%%%%%%%%%%%%%%%%%%
\subsection{Left boundary}
\label{sec:left_boundary}

Looking at Fig.~\ref{fig:Saha}, we can infer that the optical depth of the $6100\angstrom$ transition is always larger than the optical depth of the $5750\angstrom$ transition. In LTE this is always true given the values of the oscillator strengths and the higher energy of the lower level of the $5750\angstrom$ transitions. Using the parameters in Table~\ref{table:osc_strengths} and ignoring the correction for stimulated emission, we derive for LTE: 
%%%%%%%%%%%%%%%%%%%%%%%%%%%%%%%%%%%%%%%%%%%%%%%%%%
\begin{equation}
\frac{\tau_{5750}}{\tau_{6100}} \sim \frac{f_{\lambda5957}~g_{\lambda5957}+f_{\lambda5978}~g_{\lambda5978}}{f_{\lambda6371}~g_{\lambda6371}+f_{\lambda6347}~g_{\lambda6347}} e^{-\frac{\Delta E}{kT_R}} \lesssim e^{-\frac{2eV}{kT_R}} < 1
\end{equation}
%%%%%%%%%%%%%%%%%%%%%%%%%%%%%%%%%%%%%%%%%%%%%%%%%%
%%%%%%%%%%%%%%%%%%%%%%%%%%%%%%%%%%%%%%%%%%%%%%%%%%

We verified that this continues to hold after applying \textsc{Tardis}'s NLTE corrections in the following relevant range of parameters: $0.005<W<0.5$, $5\times10^{-17}$\gcm$<\rho<1\times10^{-12}$\gcm, and $5,000\,\textup{K}<T<20,000\rm\,K$. Since these hold throughout the ejecta outside the photosphere, the pEW of the $6100\angstrom$ feature will always be larger or equal to the pEW of the $5750\angstrom$ feature. This means that the $6100\angstrom$ pEW will always be larger than the $5750\angstrom$ pEW. Thus, the left boundary of the Branch plot is constrained by a 1:1 line (shown in Fig.~\ref{fig:Branch}).

The models closest to the 1:1 line will be ones with a steep Si density cutoff at low velocity. When this happens, both features begin at the photosphere with high attenuation and both stop at similar velocities. This allows the pEW of the $5750\angstrom$ feature to approach the pEW of the $6100\angstrom$ feature (see \S\ref{sec:limit_vshell}).

Fig.~\ref{fig:Density_Models} shows Si density profiles of various models. The examples given of the SCH and DDC models resemble synthetic models with $v_0=1,200$~km/s and $v_0=2,000$~km/s. According to Fig.~\ref{fig:Branch}, they would tend to the right side of the Branch plot\deleted{ -}\added{,} and in the left panel of Fig.~\ref{fig:Branch2} we see that this is indeed the case. Again in Fig.~\ref{fig:Density_Models}, we see the Si density profile of a head-on collision model (\texttt{M06\_M05\_11}) that drops steeply at $\sim12,000$~km/s. The right panel of Fig.~\ref{fig:Branch2} shows the same model on the left side of the Branch plot.

Another demonstration of this can be seen using the synthetic model in Fig.~\ref{fig:Branch}. In the left panel, the models' density decreases exponentially, and the top-left region of the Branch plot remains out of reach. In the right panel, the density is cut abruptly at some $v_{\max}$, allowing the models to cover this part of the Branch plot as well.

%%%%%%%%%%%%%%%%%%%%%%%%%%%%%%%%%%%%%%%%%%%%%%%%%%

\subsection{Right boundary}

A possible limit on the $6100\angstrom$ pEW could be interaction with the $5750\angstrom$ feature. Looking at the bottom left panel of Fig.~\ref{fig:feature_formation}, we see that the P-Cygni profile increases the flux red-ward of the rest-frame interaction wavelength (although the toy model exaggerates this effect due to lack of electron scattering). The velocity difference between the two features is only $\sim18,000$~km/s and thus interaction is possible. While in the head-on collision and synthetic models we did observe interaction and merging between features, in the observed sample the features appear to almost always be separated, so this effect is not understood to have a major effect on the shape of the right boundary. 

A better explanation is that the limit on the $6100\angstrom$ pEW arises from a physical limit on the maximum velocity of Si with $\tau_s>1$ for the $6100\angstrom$ absorption line. This is supported by the toy model presented in \S\ref{sec:toy_model}. The synthetic models of \S\ref{sec:synthetic} show that this can be translated into a condition on the Si density profile in the ejecta \deleted{-}\added{--} Fig.~\ref{fig:Branch} shows that the $6100\angstrom$ pEW is determined by the extent of the Si density profile.

However, looking at the left panel of Fig.~\ref{fig:Branch}, we see that the analytic e-folding velocity $v_0\approx2,800$~km/s discussed in \S\ref{sec:vary_v0} for an exponential density profile would result in $6100\angstrom$ pEWs larger than the observed ones. We deduce from this that the Si density drops faster than the overall ejecta density in most SNe. This is indeed the case for all explosion models presented here (see representative examples in Fig.~\ref{fig:Density_Models} for which the Si density profiles are all steeper than an exponent with $v_0=2000\, \rm km/s$, shown in a dashed-dotted gray line). 

%%%%%%%%%%%%%%%%%%%%%%%%%%%%%%%%%%%%%%%%%%%%%%%%%%
%%%%%%%%%%%%%%%%%%%%%%%%%%%%%%%%%%%%%%%%%%%%%%%%%%

\section{Discussion}

The observed Branch plot distribution is 2-dimensional, implying that 1-D models with a single variable parameter will not be able to reproduce it. This is demonstrated using the DDC and SCH models in the left panel of Fig.~\ref{fig:Branch2}.
Thus, the width of the Branch plot supports the claim that apart from luminosity, other physical parameters affect the spectra of SNe Ia \citep[e.g.][]{Hatano2000}. 
Asymmetric explosion models may provide this additional degree of freedom by allowing uncorrelated variation of the luminosity and the Si density profile based on the selection of viewing angle.

In this paper we focus on the head-on collision model \citep[e.g.][]{Rosswog2009, Raskin2010}. This model has shown promising results in terms of explaining the detonation mechanism and accounting for the observed range of $^{56}\rm Ni$ yields \citep[][]{Kushnir2013}, reproducing the observed gamma-ray escape time \citep{Kushnir2013,Nahliel} and explaining nebular spectra bi-modal emission features \citep{Kushnir2015,Vallely2020}, while facing a growing challenge in accounting for the observed rate of SNe Ia \citep[e.g.][]{Klein2017, Haim2018, Toonen2018, Hamers2018, Hallakoun2019}. 
The full coverage of the Branch plot distribution, shown in the right panel of Fig.~\ref{fig:Branch2}, provides further support for the collision model. We note that the resulting distribution does not reproduce the observed density of events across the plot, however such a comparison is premature, given that observational biases were not corrected for and 3-D ejecta spanning the range of non-zero impact parameters are not available. 
Other characteristics that have not been explored in this paper include the time-evolution and the velocities of the \ion{Si}{II} features and the distribution of $^{56}\rm Ni$ mass on the Branch plot. 

Other asymmetric models that produce ejecta with a range of Si density profiles uncorrelated with luminosity may also cover the Branch plot distribution. As a recent example, in \citet{Townsley2019} a 2-D hydrodynamical simulation of a $1\,\textup{M}_\odot$ WD double-detonation model is shown to result in an off-center detonation that produces a SN Ia that is normal in its brightness and spectra, with significant variation in the Si density profile as a function of viewing angle. Further study is required to check the Branch plot distribution of a family of these models that reproduce the relevant range of $^{56}$Ni yields.

Asymmetry in ejecta may offer an explanation for additional characteristics of SNe Ia, such as the variation in Si velocity gradients, bi-modal and shifted nebular spectral lines and the presence of high velocity features \citep[e.g.][]{Maeda2010,Blondin2011,Childress2014,Kushnir2015,Dong2018,Maguire2018}.
However, there are observational constraints that limit the possible degree of asymmetry of SNe Ia ejecta. The typically low observed continuum polarization is an important example \citep[e.g.][]{WangWheeler2008, Bulla2016, Bulla2016.2}. 
Further study is required to verify that an asymmetric origin for the spectral diversity of SNe Ia is consistent with other observed aspects. In particular, modeling of the polarization associated with the collision model is needed and is beyond the scope of this paper. 

Several assumptions and approximations limit the quantitative accuracy of our results. Calculation of the pEWs of the \ion{Si}{II} features requires a self-consistent solution of the radiation transfer problem, coupled to the solution at each location of the ionization balance and the level excitation equilibrium, which deviate from local-thermal-equilibrium (LTE). We use \textsc{Tardis}, which adopts crude approximations for the radiation field and ionization balance, affecting the expected accuracy of the simulations. In addition, the radiative transfer analysis of the 2-D collision models is performed by creating 1-D models with density and abundance sampled along sections of the 2-D ejecta from multiple viewing angles. This is not equivalent to \mbox{2-D} radiation transfer. However, the shape of an absorption feature depends mainly on the composition of material in the line of sight. Thus, we believe the qualitative analysis presented here is correct. This can be verified in the future with improved NLTE treatment and 3-D simulations.

Another result, unrelated to the symmetry of SNe Ia ejecta, stems from our analysis of the level populations relevant for the creation of the \ion{Si}{II} features (Fig.~\ref{fig:Saha}). These predict a peak of the population of the lower level of the $5750\angstrom$ transition at a photospheric temperature of $T\sim8000\rm\,K$. In our simulated models, this leads to an upper limit on the $5750\angstrom$ pEW at low luminosities. This saturation effect also manifests in the other models mentioned in this paper, which are based on independent radiative transfer models (see left panel of Fig.~\ref{fig:Branch2}). Currently, plots of the \ion{Si}{II} features' pEW vs. luminosity tracers such as $\Delta \rm m_{15}(\rm B)$ do not show a significant saturation effect (see e.g. Fig. 17 in \citealt{CSP2013}). If such an effect is identified in future observations of lower luminosity SNe it may help calibrate the photospheric temperature and provide a useful tool for constraining progenitor models.

%%%%%%%%%%%%%%%%%%%%%%%%%%%%%%%%%%%%%%%%%%%%%%%%%%
%%%%%%%%%%%%%%%%%%%%%%%%%%%%%%%%%%%%%%%%%%%%%%%%%%

\section{Summary}

As part of the ongoing effort to identify the nature of the progenitor system of \deleted{t}\added{T}ype Ia supernovae, we study the spectra of these objects at maximum bolometric luminosity. We focus on the \ion{Si}{II} $6100\angstrom$ and $5750\angstrom$ features and use the Branch plot, a 2-D plot of the pEW distribution \citep{Branch2006}, as a tool to test the validity of different models.

The main result of this paper is presented in Fig.~\ref{fig:Branch2}, where the distribution of \ion{Si}{II} pEWs for simulated hydrodynamical explosion models is compared to observations. \deleted{In the left panel, the results of the spherically symmetric central detonations of sub-Chandrasekhar WDs (SCH) and delayed detonation Chandrasekhar models (DDC) are shown, using both pEWs extracted from spectra calculated by these authors and our results of Tardis simulations of the same ejecta (see). In the right panel, Tardis simulations of ejecta profiles along different viewing angles of the asymmetric WD head-on collision model are shown (see.)}The 1-D \added{SCH and DDC} models fail to reach most of the observed range of pEWs, while the head-on collision model shows almost full coverage of the observed distribution (for sample spectra see Appendix~\ref{sec:Spectra}). As shown in this paper, the success of the head-on collision model in reproducing the observed distribution on the Branch plot is a result of its asymmetry, which allows for a significant range of Si density profiles along different viewing angles (Fig.~\ref{fig:Density_Models}), coupled but uncorrelated with a range of $^{56}$Ni yields that cover the observed range of SNe Ia luminosity. 
	
% In the left panel, the results of the spherically symmetric central detonations of sub-Chandrasekhar WDs (SCH) and delayed detonation Chandrasekhar models (DDC) \citep{Blondin2013,Blondin2017} are shown, using both pEWs extracted from spectra calculated by these authors and our results of \textsc{Tardis} simulations of the same ejecta (see \S\ref{sec:models}). In the right panel, \textsc{Tardis} simulations of ejecta profiles along different viewing angles of the asymmetric WD head-on collision model \citep{Kushnir2013} are shown (see \S\ref{sec:collision}).	

An order-of-magnitude analysis of the formation of the \ion{Si}{II} features is performed in \S\ref{sec:formation}. \deleted{Using LTE and simple NLTE approximations, and taking typical photospheric temperature and density for SNe at maximum light, }The Si density is shown to be 10 to 100 times the required density for optical depth of unity for these features (Fig.~\ref{fig:Saha}). Given that the Si density naturally drops by more than 2 orders of magnitudes between $10,000~\rm km/s$ and $20,000~\rm km/s$, it is reasonable that \deleted{t}\added{T}ype Ia's have absorption regions that are within this range. \deleted{As shown, }The population of the relevant \ion{Si}{II} excited levels rises with decreasing temperature, explaining the dependence of the \ion{Si}{II} pEWs on SN luminosity. The population \deleted{is shown to }peaks at $T\sim8000\rm\,K$, causing a saturation effect (see \S\ref{sec:low_lumin}) and predicting a maximum pEW at low luminosity.

In an attempt to clarify the effects of geometry on the spectral features, in \S\ref{sec:toy_model} a toy model of an ejecta containing a single absorption line is simulated. The resulting spectral features are shown to reproduce the non-trivial relations between the pEW and both the fractional depth and the FWHM (\deleted{top panels of }Fig.~\ref{fig:feature_formation}). These results support our assumption that the \ion{Si}{II} features can be analyzed as being due to a single absorption line, and show that the pEW is a good tracer for the extent of the absorption region\deleted{ (bottom right panel of Fig.~4)}.

In \S\ref{sec:synthetic} \textsc{Tardis} is applied to simplified synthetic ejecta with exponentially declining density models. The observed Branch plot distribution is reproduced by varying two factors: the luminosity of the SN and the Si density profile of the ejecta (Fig.~\ref{fig:Branch}). Specifically, introducing a cutoff to the Si density profile at low $v_{\max}$ leads to \ion{Si}{II} features located on the elusive top left part of the Branch plot.\deleted{ The population peak at $T\sim8000\rm\,K$ (see Fig.~3) is shown to manifest as a saturation effect for low luminosity SNe.}

Realizing the importance of the Si density profile, we numerically find an approximate Si density threshold, predicting the extent of the absorption region and the pEW of the $6100\angstrom$ feature. The use of this threshold on sample ejecta is demonstrated in Fig.~\ref{fig:Velocity}, showing a clear correlation between the effective "maximum Si velocity" in the ejecta and the $6100\angstrom$ pEW for low luminosity head-on collision and synthetic models.

Based on the above results, the bounds of the Branch plot are explained: the top boundary represents either a limit on the lowest luminosity of SNe Ia events or a limit on the maximum optical depth from atomic physics \deleted{(}due to the saturation effect\deleted{,} \added{(}see \S\ref{sec:top_boundary}). The left boundary is constrained by a 1:1 line from atomic physics (shown in Fig.~\ref{fig:Branch}), and it seems that for low-luminosity events to approach it requires a steeply falling Si density profile (\S\ref{sec:left_boundary}). The right boundary represents an upper limit on the velocity of Si in the ejecta.

%%%%%%%%%%%%%%%%%%%%%%%%%%%%%%%%%%%%%%%%%%%%%%%%%%
%%%%%%%%%%%%%%%%%%%%%%%%%%%%%%%%%%%%%%%%%%%%%%%%%%

\section*{Acknowledgements}

We thank Doron Kushnir and St\'ephane Blondin for sharing model ejecta and spectra.
We thank Nahliel Wygoda for sharing light-curve data.
We thank Eli Waxman, Avishay Gal-Yam and Doron Kushnir for useful discussions.
We thank Wolfgang Kerzendorf and other \textsc{Tardis} contributors for their support.
\added{We thank the anonymous referee for helpful comments.}
This work was supported by the Beracha foundation and the Minerva foundation with funding from the Federal German Ministry for Education and Research.
This research made use of \textsc{Tardis}, a community-developed software package for spectral synthesis in supernovae \added{\mbox{\citep{Tardis, kerzendorf_wolfgang_2019_2590539}}}. The development of \textsc{Tardis} received support from the Google Summer of Code initiative and from ESA's Summer of Code in Space program. \textsc{Tardis} makes extensive use of Astropy and PyNE.


\bibliographystyle{mnras}
\bibliography{bib_file} 

\begin{thebibliography}{}
\makeatletter
\relax
\def\mn@urlcharsother{\let\do\@makeother \do\$\do\&\do\#\do\^\do\_\do\%\do\~}
\def\mn@doi{\begingroup\mn@urlcharsother \@ifnextchar [ {\mn@doi@}
  {\mn@doi@[]}}
\def\mn@doi@[#1]#2{\def\@tempa{#1}\ifx\@tempa\@empty \href
  {http://dx.doi.org/#2} {doi:#2}\else \href {http://dx.doi.org/#2} {#1}\fi
  \endgroup}
\def\mn@eprint#1#2{\mn@eprint@#1:#2::\@nil}
\def\mn@eprint@arXiv#1{\href {http://arxiv.org/abs/#1} {{\tt arXiv:#1}}}
\def\mn@eprint@dblp#1{\href {http://dblp.uni-trier.de/rec/bibtex/#1.xml}
  {dblp:#1}}
\def\mn@eprint@#1:#2:#3:#4\@nil{\def\@tempa {#1}\def\@tempb {#2}\def\@tempc
  {#3}\ifx \@tempc \@empty \let \@tempc \@tempb \let \@tempb \@tempa \fi \ifx
  \@tempb \@empty \def\@tempb {arXiv}\fi \@ifundefined
  {mn@eprint@\@tempb}{\@tempb:\@tempc}{\expandafter \expandafter \csname
  mn@eprint@\@tempb\endcsname \expandafter{\@tempc}}}

\bibitem[\protect\citeauthoryear{{Benetti} et~al.,}{{Benetti}
  et~al.}{2005}]{Benetti2005}
{Benetti} S.,  et~al., 2005, \mn@doi [\apj] {10.1086/428608}, \href
  {https://ui.adsabs.harvard.edu/abs/2005ApJ...623.1011B} {623, 1011}

\bibitem[\protect\citeauthoryear{{Blondin} \& {Tonry}}{{Blondin} \&
  {Tonry}}{2007}]{SNID2007}
{Blondin} S.,  {Tonry} J.~L.,  2007, \mn@doi [\apj] {10.1086/520494}, \href
  {https://ui.adsabs.harvard.edu/\#abs/2007ApJ...666.1024B} {666, 1024}

\bibitem[\protect\citeauthoryear{{Blondin}, {Kasen}, {R{\"o}pke}, {Kirshner}
  \& {Mandel}}{{Blondin} et~al.}{2011}]{Blondin2011}
{Blondin} S.,  {Kasen} D.,  {R{\"o}pke} F.~K.,  {Kirshner} R.~P.,   {Mandel}
  K.~S.,  2011, \mn@doi [\mnras] {10.1111/j.1365-2966.2011.19345.x}, \href
  {https://ui.adsabs.harvard.edu/\#abs/2011MNRAS.417.1280B} {417, 1280}

\bibitem[\protect\citeauthoryear{{Blondin} et~al.,}{{Blondin}
  et~al.}{2012}]{CfA2012}
{Blondin} S.,  et~al., 2012, \mn@doi [\aj] {10.1088/0004-6256/143/5/126}, \href
  {https://ui.adsabs.harvard.edu/#abs/2012AJ....143..126B} {143, 126}

\bibitem[\protect\citeauthoryear{{Blondin}, {Dessart}, {Hillier}  \&
  {Khokhlov}}{{Blondin} et~al.}{2013}]{Blondin2013}
{Blondin} S.,  {Dessart} L.,  {Hillier} D.~J.,   {Khokhlov} A.~M.,  2013,
  \mn@doi [\mnras] {10.1093/mnras/sts484}, \href
  {https://ui.adsabs.harvard.edu/#abs/2013MNRAS.429.2127B} {429, 2127}

\bibitem[\protect\citeauthoryear{{Blondin}, {Dessart}, {Hillier}  \&
  {Khokhlov}}{{Blondin} et~al.}{2017}]{Blondin2017}
{Blondin} S.,  {Dessart} L.,  {Hillier} D.~J.,   {Khokhlov} A.~M.,  2017,
  \mn@doi [\mnras] {10.1093/mnras/stw2492}, \href
  {https://ui.adsabs.harvard.edu/#abs/2017MNRAS.470..157B} {470, 157}

\bibitem[\protect\citeauthoryear{{Branch} et~al.,}{{Branch}
  et~al.}{2006}]{Branch2006}
{Branch} D.,  et~al., 2006, \mn@doi [Publications of the Astronomical Society
  of the Pacific] {10.1086/502778}, \href
  {https://ui.adsabs.harvard.edu/\#abs/2006PASP..118..560B} {118, 560}

\bibitem[\protect\citeauthoryear{{Branch}, {Chau Dang}  \& {Baron}}{{Branch}
  et~al.}{2009}]{Branch2009}
{Branch} D.,  {Chau Dang} L.,   {Baron} E.,  2009, \mn@doi [Publications of the
  Astronomical Society of the Pacific] {10.1086/597788}, \href
  {https://ui.adsabs.harvard.edu/#abs/2009PASP..121..238B} {121, 238}

\bibitem[\protect\citeauthoryear{{Bulla}, {Sim}, {Pakmor}, {Kromer},
  {Taubenberger}, {R{\"o}pke}, {Hillebrandt}  \& {Seitenzahl}}{{Bulla}
  et~al.}{2016a}]{Bulla2016}
{Bulla} M.,  {Sim} S.~A.,  {Pakmor} R.,  {Kromer} M.,  {Taubenberger} S.,
  {R{\"o}pke} F.~K.,  {Hillebrandt} W.,   {Seitenzahl} I.~R.,  2016a, \mn@doi
  [\mnras] {10.1093/mnras/stv2402}, \href
  {https://ui.adsabs.harvard.edu/abs/2016MNRAS.455.1060B} {455, 1060}

\bibitem[\protect\citeauthoryear{{Bulla} et~al.,}{{Bulla}
  et~al.}{2016b}]{Bulla2016.2}
{Bulla} M.,  et~al., 2016b, \mn@doi [\mnras] {10.1093/mnras/stw1733}, \href
  {https://ui.adsabs.harvard.edu/abs/2016MNRAS.462.1039B} {462, 1039}

\bibitem[\protect\citeauthoryear{{Childress}, {Filippenko}, {Ganeshalingam}  \&
  {Schmidt}}{{Childress} et~al.}{2014}]{Childress2014}
{Childress} M.~J.,  {Filippenko} A.~V.,  {Ganeshalingam} M.,   {Schmidt} B.~P.,
   2014, \mn@doi [\mnras] {10.1093/mnras/stt1892}, \href
  {https://ui.adsabs.harvard.edu/abs/2014MNRAS.437..338C} {437, 338}

\bibitem[\protect\citeauthoryear{{Dong}, {Katz}, {Kushnir}  \& {Prieto}}{{Dong}
  et~al.}{2015}]{Kushnir2015}
{Dong} S.,  {Katz} B.,  {Kushnir} D.,   {Prieto} J.~L.,  2015, \mn@doi [\mnras]
  {10.1093/mnrasl/slv129}, \href
  {https://ui.adsabs.harvard.edu/\#abs/2015MNRAS.454L..61D} {454, L61}

\bibitem[\protect\citeauthoryear{{Dong} et~al.,}{{Dong}
  et~al.}{2018}]{Dong2018}
{Dong} S.,  et~al., 2018, \mn@doi [\mnras] {10.1093/mnrasl/sly098}, \href
  {https://ui.adsabs.harvard.edu/abs/2018MNRAS.479L..70D} {479, L70}

\bibitem[\protect\citeauthoryear{{Folatelli} et~al.,}{{Folatelli}
  et~al.}{2013}]{CSP2013}
{Folatelli} G.,  et~al., 2013, \mn@doi [\apj] {10.1088/0004-637X/773/1/53},
  \href {https://ui.adsabs.harvard.edu/#abs/2013ApJ...773...53F} {773, 53}

\bibitem[\protect\citeauthoryear{{Hachinger}, {Mazzali}, {Tanaka},
  {Hillebrandt}  \& {Benetti}}{{Hachinger} et~al.}{2008}]{Hachinger2008}
{Hachinger} S.,  {Mazzali} P.~A.,  {Tanaka} M.,  {Hillebrandt} W.,   {Benetti}
  S.,  2008, \mn@doi [\mnras] {10.1111/j.1365-2966.2008.13645.x}, \href
  {https://ui.adsabs.harvard.edu/#abs/2008MNRAS.389.1087H} {389, 1087}

\bibitem[\protect\citeauthoryear{{Haim} \& {Katz}}{{Haim} \&
  {Katz}}{2018}]{Haim2018}
{Haim} N.,  {Katz} B.,  2018, \mn@doi [\mnras] {10.1093/mnras/sty1588}, \href
  {https://ui.adsabs.harvard.edu/abs/2018MNRAS.479.3155H} {479, 3155}

\bibitem[\protect\citeauthoryear{{Hallakoun} \& {Maoz}}{{Hallakoun} \&
  {Maoz}}{2019}]{Hallakoun2019}
{Hallakoun} N.,  {Maoz} D.,  2019, \mn@doi [\mnras] {10.1093/mnras/stz2535},
  \href {https://ui.adsabs.harvard.edu/abs/2019MNRAS.490..657H} {490, 657}

\bibitem[\protect\citeauthoryear{{Hamers}}{{Hamers}}{2018}]{Hamers2018}
{Hamers} A.~S.,  2018, \mn@doi [\mnras] {10.1093/mnras/sty985}, \href
  {https://ui.adsabs.harvard.edu/abs/2018MNRAS.478..620H} {478, 620}

\bibitem[\protect\citeauthoryear{{Hatano}, {Branch}, {Lentz}, {Baron},
  {Filippenko}  \& {Garnavich}}{{Hatano} et~al.}{2000}]{Hatano2000}
{Hatano} K.,  {Branch} D.,  {Lentz} E.~J.,  {Baron} E.,  {Filippenko} A.~V.,
  {Garnavich} P.~M.,  2000, \mn@doi [\apjl] {10.1086/318169}, \href
  {https://ui.adsabs.harvard.edu/abs/2000ApJ...543L..49H} {543, L49}

\bibitem[\protect\citeauthoryear{Heringer, van Kerkwijk, Sim  \&
  Kerzendorf}{Heringer et~al.}{2017}]{Heringer2017}
Heringer E.,  van Kerkwijk M.~H.,  Sim S.~A.,   Kerzendorf W.~E.,  2017,
  \mn@doi [Astrophys. J.] {10.3847/1538-4357/aa8309}, 846, 15

\bibitem[\protect\citeauthoryear{{Hillier} \& {Miller}}{{Hillier} \&
  {Miller}}{1998}]{CMFGEN}
{Hillier} D.~J.,  {Miller} D.~L.,  1998, \mn@doi [\apj] {10.1086/305350}, \href
  {https://ui.adsabs.harvard.edu/abs/1998ApJ...496..407H} {496, 407}

\bibitem[\protect\citeauthoryear{{Jeffery}}{{Jeffery}}{1999}]{Jeffery1999}
{Jeffery} D.~J.,  1999, arXiv e-prints, \href
  {https://ui.adsabs.harvard.edu/\#abs/1999astro.ph..7015J} {pp
  astro--ph/9907015}

\bibitem[\protect\citeauthoryear{{Kerzendorf} \& {Sim}}{{Kerzendorf} \&
  {Sim}}{2014}]{Tardis}
{Kerzendorf} W.~E.,  {Sim} S.~A.,  2014, \mn@doi [\mnras]
  {10.1093/mnras/stu055}, \href
  {http://adsabs.harvard.edu/abs/2014MNRAS.440..387K} {440, 387}

\bibitem[\protect\citeauthoryear{Kerzendorf et~al.,}{Kerzendorf
  et~al.}{2019}]{kerzendorf_wolfgang_2019_2590539}
Kerzendorf W.,  et~al., 2019, tardis-sn/tardis: TARDIS v3.0 alpha2,
  \mn@doi{10.5281/zenodo.2590539}, \url
  {https://doi.org/10.5281/zenodo.2590539}

\bibitem[\protect\citeauthoryear{{Khokhlov}}{{Khokhlov}}{1991}]{Khokhlov1991}
{Khokhlov} A.~M.,  1991, \aap, \href
  {https://ui.adsabs.harvard.edu/abs/1991A&A...245..114K} {245, 114}

\bibitem[\protect\citeauthoryear{{Klein} \& {Katz}}{{Klein} \&
  {Katz}}{2017}]{Klein2017}
{Klein} Y.~Y.,  {Katz} B.,  2017, \mn@doi [\mnras] {10.1093/mnrasl/slw207},
  \href {https://ui.adsabs.harvard.edu/abs/2017MNRAS.465L..44K} {465, L44}

\bibitem[\protect\citeauthoryear{Kramida, {Yu.~Ralchenko}, Reader  \& {and NIST
  ASD Team}}{Kramida et~al.}{2019}]{NIST_ASD}
Kramida A.,  {Yu.~Ralchenko} Reader J.,   {and NIST ASD Team} 2019, {NIST
  Atomic Spectra Database (ver. 5.7.1), [Online]. Available:
  {\tt{https://physics.nist.gov/asd}} [2019, December 5]. National Institute of
  Standards and Technology, Gaithersburg, MD.}

\bibitem[\protect\citeauthoryear{{Kurucz} \& {Bell}}{{Kurucz} \&
  {Bell}}{1995}]{Kurucz}
{Kurucz} R.~L.,  {Bell} B.,  1995, {Atomic line list}

\bibitem[\protect\citeauthoryear{{Kushnir}, {Katz}, {Dong}, {Livne}  \&
  {Fern{\'a}ndez}}{{Kushnir} et~al.}{2013}]{Kushnir2013}
{Kushnir} D.,  {Katz} B.,  {Dong} S.,  {Livne} E.,   {Fern{\'a}ndez} R.,  2013,
  \mn@doi [\apj] {10.1088/2041-8205/778/2/L37}, \href
  {https://ui.adsabs.harvard.edu/#abs/2013ApJ...778L..37K} {778, L37}

\bibitem[\protect\citeauthoryear{{Livio} \& {Mazzali}}{{Livio} \&
  {Mazzali}}{2018}]{Livio2018}
{Livio} M.,  {Mazzali} P.,  2018, \mn@doi [\physrep]
  {10.1016/j.physrep.2018.02.002}, \href
  {https://ui.adsabs.harvard.edu/\#abs/2018PhR...736....1L} {736, 1}

\bibitem[\protect\citeauthoryear{{Maeda} et~al.,}{{Maeda}
  et~al.}{2010}]{Maeda2010}
{Maeda} K.,  et~al., 2010, \mn@doi [\nat] {10.1038/nature09122}, \href
  {https://ui.adsabs.harvard.edu/\#abs/2010Natur.466...82M} {466, 82}

\bibitem[\protect\citeauthoryear{{Maguire} et~al.,}{{Maguire}
  et~al.}{2018}]{Maguire2018}
{Maguire} K.,  et~al., 2018, \mn@doi [\mnras] {10.1093/mnras/sty820}, \href
  {https://ui.adsabs.harvard.edu/abs/2018MNRAS.477.3567M} {477, 3567}

\bibitem[\protect\citeauthoryear{{Maoz}, {Mannucci}  \& {Nelemans}}{{Maoz}
  et~al.}{2014}]{Maoz2014}
{Maoz} D.,  {Mannucci} F.,   {Nelemans} G.,  2014, \mn@doi [Annual Review of
  Astronomy and Astrophysics] {10.1146/annurev-astro-082812-141031}, \href
  {https://ui.adsabs.harvard.edu/\#abs/2014ARA&A..52..107M} {52, 107}

\bibitem[\protect\citeauthoryear{{Mazzali} \& {Lucy}}{{Mazzali} \&
  {Lucy}}{1993}]{ML93}
{Mazzali} P.~A.,  {Lucy} L.~B.,  1993, \aap, \href
  {https://ui.adsabs.harvard.edu/abs/1993A&A...279..447M} {279, 447}

\bibitem[\protect\citeauthoryear{{Nomoto}}{{Nomoto}}{1982}]{Nomoto1982}
{Nomoto} K.,  1982, \mn@doi [\apj] {10.1086/159682}, \href
  {https://ui.adsabs.harvard.edu/abs/1982ApJ...253..798N} {253, 798}

\bibitem[\protect\citeauthoryear{{Nomoto}, {Thielemann}  \& {Yokoi}}{{Nomoto}
  et~al.}{1984}]{Nomoto1984}
{Nomoto} K.,  {Thielemann} F.~K.,   {Yokoi} K.,  1984, \mn@doi [\apj]
  {10.1086/162639}, \href
  {https://ui.adsabs.harvard.edu/abs/1984ApJ...286..644N} {286, 644}

\bibitem[\protect\citeauthoryear{{Nugent}, {Phillips}, {Baron}, {Branch}  \&
  {Hauschildt}}{{Nugent} et~al.}{1995}]{Nugent1995}
{Nugent} P.,  {Phillips} M.,  {Baron} E.,  {Branch} D.,   {Hauschildt} P.,
  1995, \mn@doi [\apj] {10.1086/309846}, \href
  {https://ui.adsabs.harvard.edu/#abs/1995ApJ...455L.147N} {455, L147}

\bibitem[\protect\citeauthoryear{{Raskin}, {Scannapieco}, {Rockefeller},
  {Fryer}, {Diehl}  \& {Timmes}}{{Raskin} et~al.}{2010}]{Raskin2010}
{Raskin} C.,  {Scannapieco} E.,  {Rockefeller} G.,  {Fryer} C.,  {Diehl} S.,
  {Timmes} F.~X.,  2010, \mn@doi [\apj] {10.1088/0004-637X/724/1/111}, \href
  {https://ui.adsabs.harvard.edu/abs/2010ApJ...724..111R} {724, 111}

\bibitem[\protect\citeauthoryear{{Rosswog}, {Kasen}, {Guillochon}  \&
  {Ramirez-Ruiz}}{{Rosswog} et~al.}{2009}]{Rosswog2009}
{Rosswog} S.,  {Kasen} D.,  {Guillochon} J.,   {Ramirez-Ruiz} E.,  2009,
  \mn@doi [\apjl] {10.1088/0004-637X/705/2/L128}, \href
  {https://ui.adsabs.harvard.edu/abs/2009ApJ...705L.128R} {705, L128}

\bibitem[\protect\citeauthoryear{{Savitzky} \& {Golay}}{{Savitzky} \&
  {Golay}}{1964}]{SavGol64}
{Savitzky} A.,  {Golay} M.~J.~E.,  1964, Analytical Chemistry, \href
  {https://ui.adsabs.harvard.edu/abs/1964AnaCh..36.1627S} {36, 1627}

\bibitem[\protect\citeauthoryear{{Silverman}, {Kong}  \&
  {Filippenko}}{{Silverman} et~al.}{2012}]{Silverman2012}
{Silverman} J.~M.,  {Kong} J.~J.,   {Filippenko} A.~V.,  2012, \mn@doi [\mnras]
  {10.1111/j.1365-2966.2012.21269.x}, \href
  {https://ui.adsabs.harvard.edu/#abs/2012MNRAS.425.1819S} {425, 1819}

\bibitem[\protect\citeauthoryear{{Sim}, {R{\"o}pke}, {Hillebrandt}, {Kromer},
  {Pakmor}, {Fink}, {Ruiter}  \& {Seitenzahl}}{{Sim} et~al.}{2010}]{Sim2010}
{Sim} S.~A.,  {R{\"o}pke} F.~K.,  {Hillebrandt} W.,  {Kromer} M.,  {Pakmor} R.,
   {Fink} M.,  {Ruiter} A.~J.,   {Seitenzahl} I.~R.,  2010, \mn@doi [\apjl]
  {10.1088/2041-8205/714/1/L52}, \href
  {https://ui.adsabs.harvard.edu/abs/2010ApJ...714L..52S} {714, L52}

\bibitem[\protect\citeauthoryear{{Soker}}{{Soker}}{2019}]{Soker2019}
{Soker} N.,  2019, arXiv e-prints, \href
  {https://ui.adsabs.harvard.edu/abs/2019arXiv191201550S} {p. arXiv:1912.01550}

\bibitem[\protect\citeauthoryear{Toonen, Perets  \& Hamers}{Toonen
  et~al.}{2018}]{Toonen2018}
Toonen S.,  Perets H.~B.,   Hamers A.~S.,  2018, \mn@doi [Astronomy &
  Astrophysics] {10.1051/0004-6361/201731874}, 610, A22

\bibitem[\protect\citeauthoryear{{Townsley}, {Miles}, {Shen}  \&
  {Kasen}}{{Townsley} et~al.}{2019}]{Townsley2019}
{Townsley} D.~M.,  {Miles} B.~J.,  {Shen} K.~J.,   {Kasen} D.,  2019, \mn@doi
  [\apjl] {10.3847/2041-8213/ab27cd}, \href
  {https://ui.adsabs.harvard.edu/abs/2019ApJ...878L..38T} {878, L38}

\bibitem[\protect\citeauthoryear{{Vallely}, {Tucker}, {Shappee}, {Brown},
  {Stanek}  \& {Kochanek}}{{Vallely} et~al.}{2020}]{Vallely2020}
{Vallely} P.~J.,  {Tucker} M.~A.,  {Shappee} B.~J.,  {Brown} J.~S.,  {Stanek}
  K.~Z.,   {Kochanek} C.~S.,  2020, \mn@doi [\mnras] {10.1093/mnras/staa003},
  \href {https://ui.adsabs.harvard.edu/abs/2020MNRAS.492.3553V} {492, 3553}

\bibitem[\protect\citeauthoryear{{Wang} \& {Wheeler}}{{Wang} \&
  {Wheeler}}{2008}]{WangWheeler2008}
{Wang} L.,  {Wheeler} J.~C.,  2008, \mn@doi [\araa]
  {10.1146/annurev.astro.46.060407.145139}, \href
  {https://ui.adsabs.harvard.edu/abs/2008ARA&A..46..433W} {46, 433}

\bibitem[\protect\citeauthoryear{{Wang} et~al.,}{{Wang}
  et~al.}{2009}]{Wang2009}
{Wang} X.,  et~al., 2009, \mn@doi [\apjl] {10.1088/0004-637X/699/2/L139}, \href
  {https://ui.adsabs.harvard.edu/abs/2009ApJ...699L.139W} {699, L139}

\bibitem[\protect\citeauthoryear{{Wilk}, {Hillier}  \& {Dessart}}{{Wilk}
  et~al.}{2018}]{Wilk2018}
{Wilk} K.~D.,  {Hillier} D.~J.,   {Dessart} L.,  2018, \mn@doi [\mnras]
  {10.1093/mnras/stx2816}, \href
  {https://ui.adsabs.harvard.edu/abs/2018MNRAS.474.3187W} {474, 3187}

\bibitem[\protect\citeauthoryear{{Wygoda}, {Elbaz}  \& {Katz}}{{Wygoda}
  et~al.}{2019a}]{Nahliel}
{Wygoda} N.,  {Elbaz} Y.,   {Katz} B.,  2019a, \mn@doi [\mnras]
  {10.1093/mnras/stz145}, \href
  {https://ui.adsabs.harvard.edu/abs/2019MNRAS.484.3941W} {484, 3941}

\bibitem[\protect\citeauthoryear{{Wygoda}, {Elbaz}  \& {Katz}}{{Wygoda}
  et~al.}{2019b}]{Wygoda.2}
{Wygoda} N.,  {Elbaz} Y.,   {Katz} B.,  2019b, \mn@doi [\mnras]
  {10.1093/mnras/stz146}, \href
  {https://ui.adsabs.harvard.edu/abs/2019MNRAS.484.3951W} {484, 3951}

\bibitem[\protect\citeauthoryear{{Yaron} \& {Gal-Yam}}{{Yaron} \&
  {Gal-Yam}}{2012}]{WiseRep}
{Yaron} O.,  {Gal-Yam} A.,  2012, \mn@doi [Publications of the Astronomical
  Society of the Pacific] {10.1086/666656}, \href
  {https://ui.adsabs.harvard.edu/#abs/2012PASP..124..668Y} {124, 668}

\makeatother
\end{thebibliography}

%%%%%%%%%%%%%%%%%%%%%%%%%%%%%%%%%%%%%%%%%%%%%%%%%%
%%%%%%%%%%%%%%%%% APPENDICES %%%%%%%%%%%%%%%%%%%%%
\appendix

%%%%%%%%%%%%%%%%%%%%%%%%%%%%%%%%%%%%%%%%%%%%%%%%%%
%%%%%%%%%%%%%%%%%%%%%%%%%%%%%%%%%%%%%%%%%%%%%%%%%%

\section{Relations between line parameters in simulated models}
\label{sec:simulated}
\begin{figure*}
	\centering
	\includegraphics[width=\columnwidth]{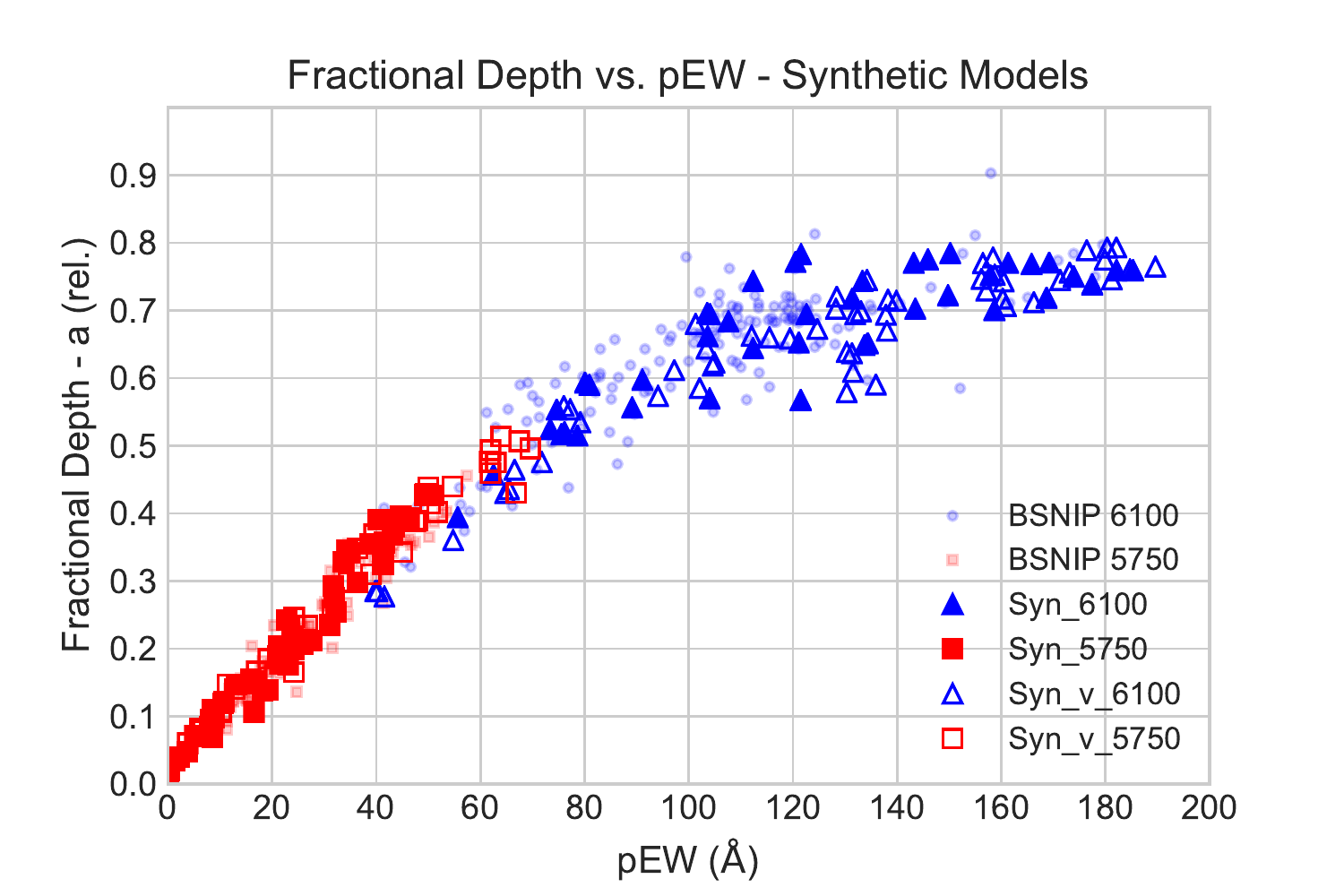}
	~
	\includegraphics[width=\columnwidth]{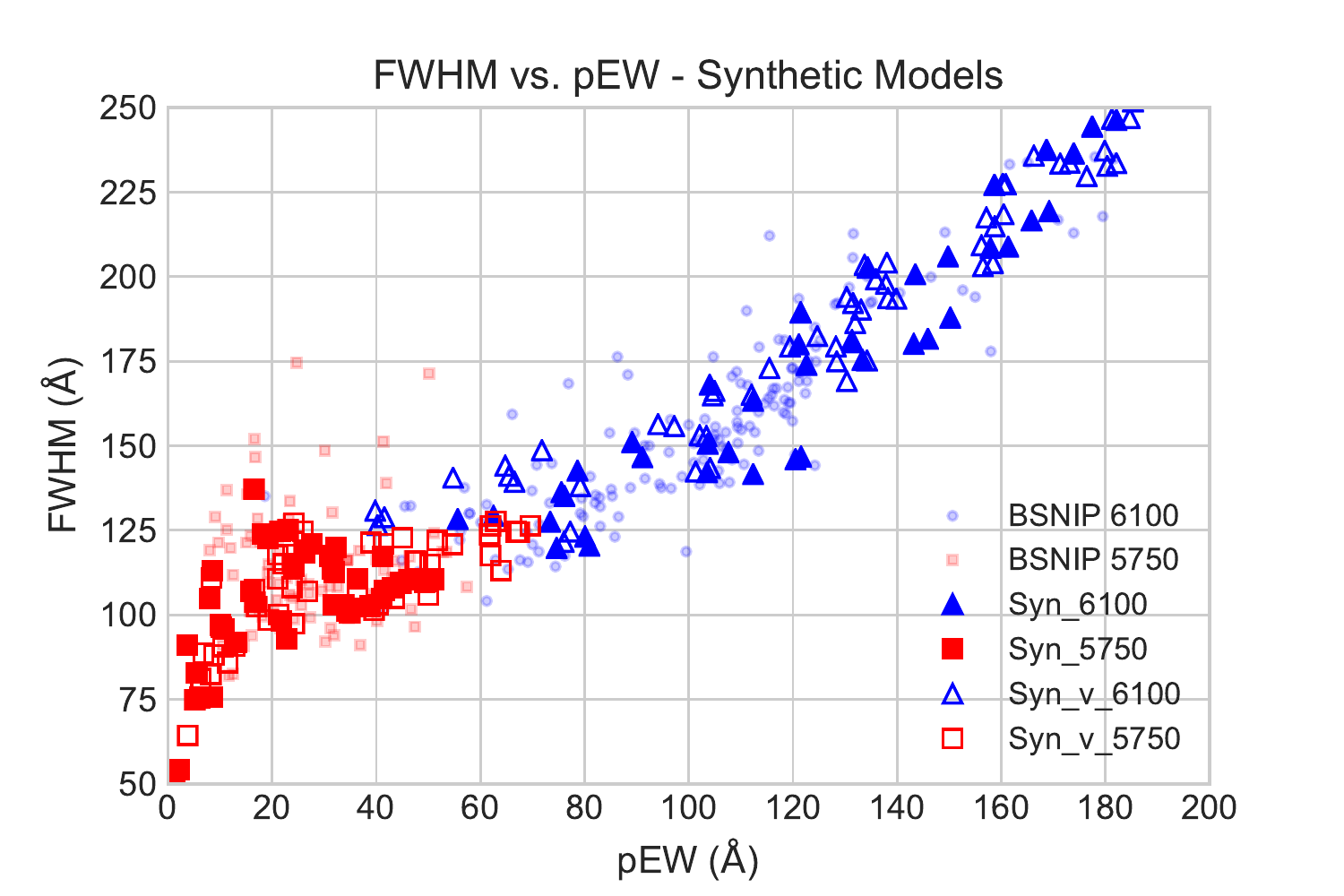}
	\hfill
	\includegraphics[width=\columnwidth]{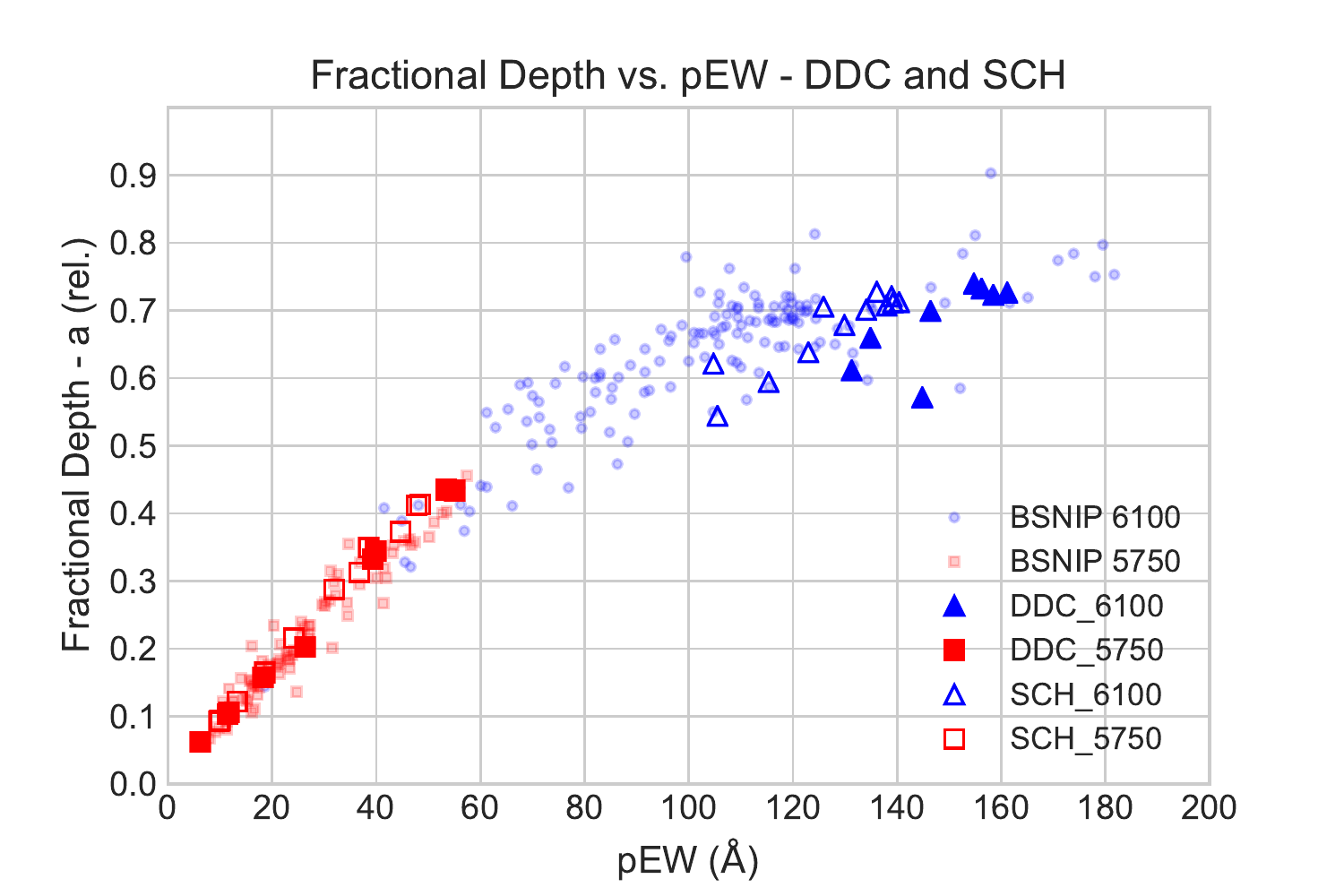}
	~ 
	\includegraphics[width=\columnwidth]{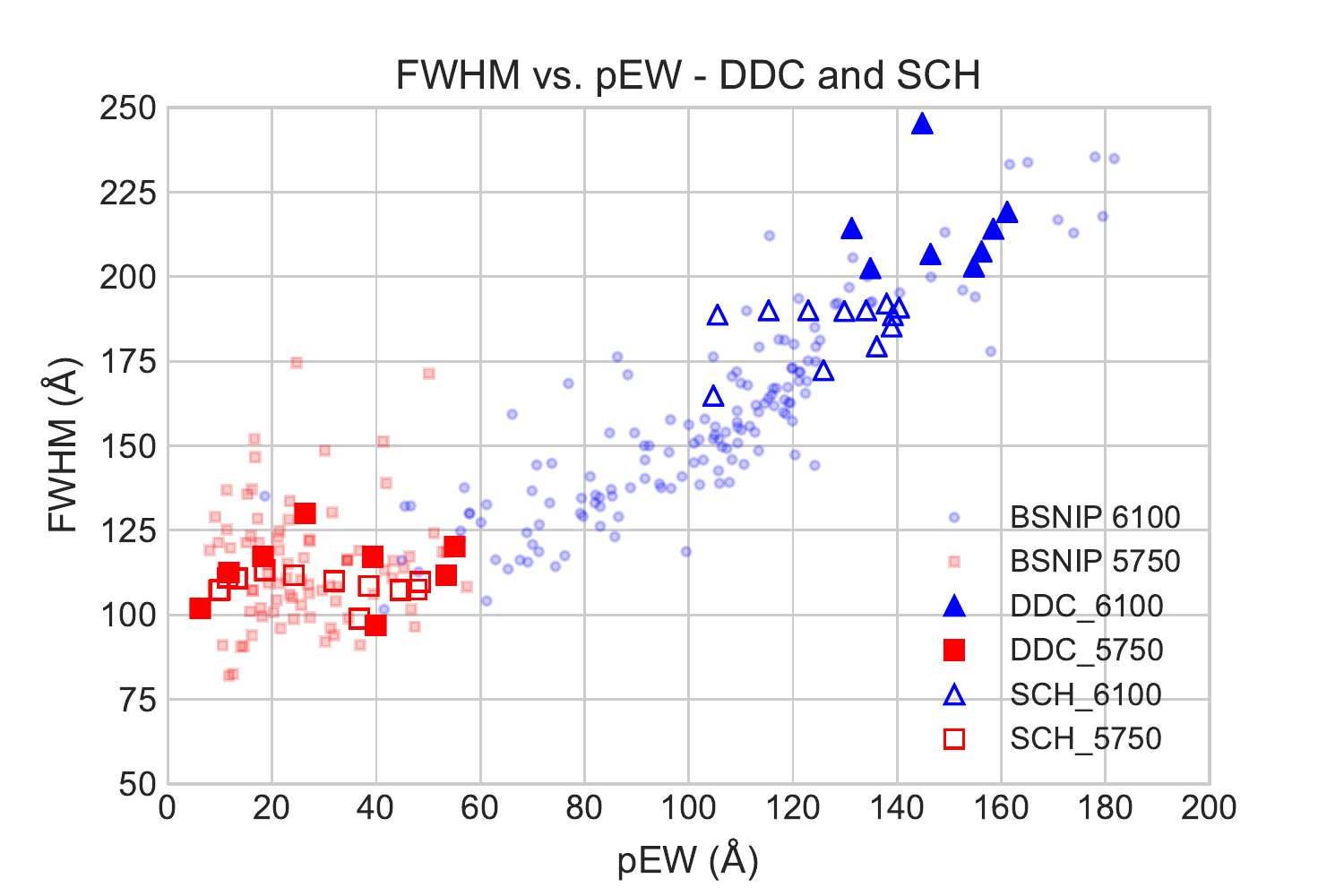}
	\hfill
	\includegraphics[width=\columnwidth]{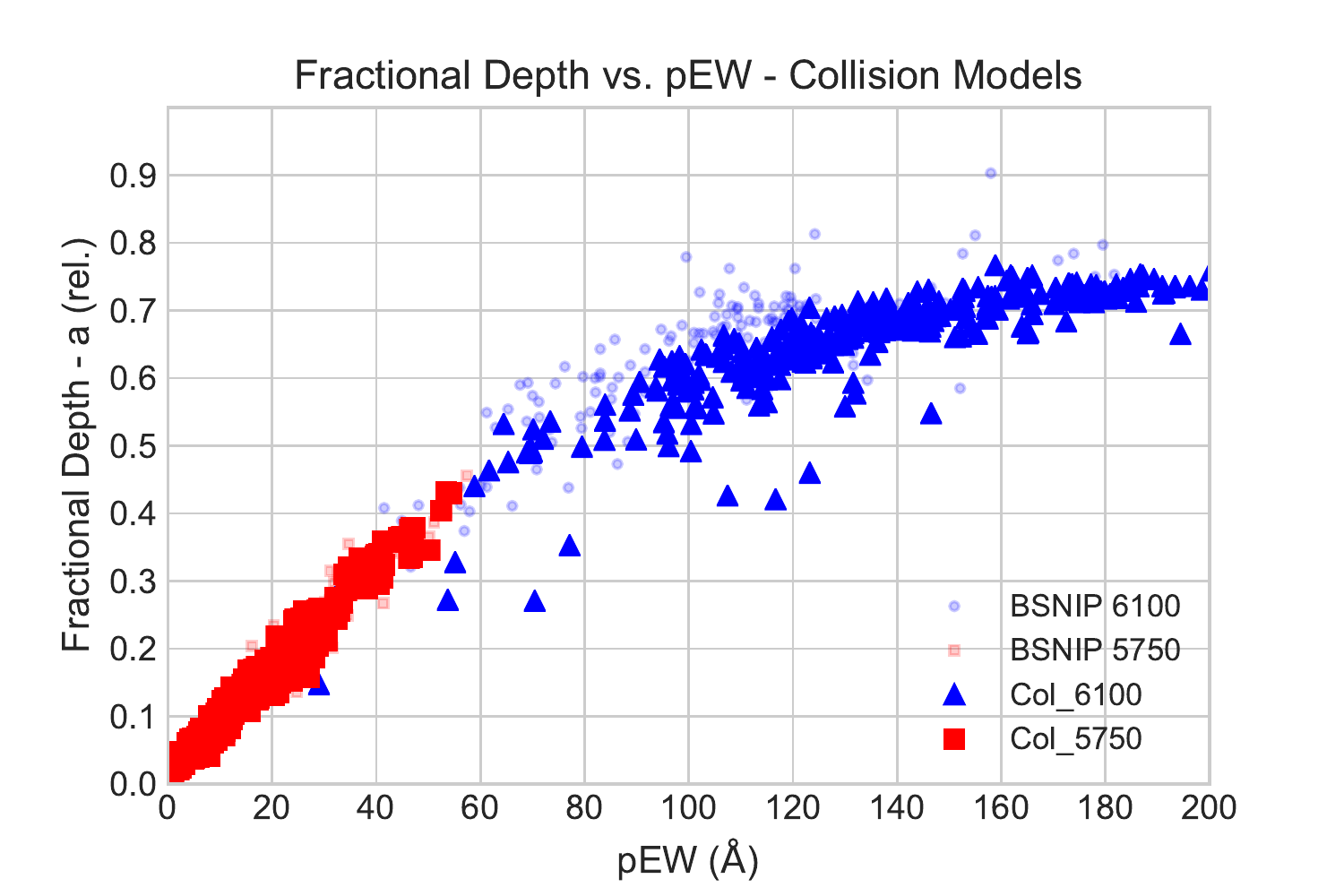}
	~ 
	\includegraphics[width=\columnwidth]{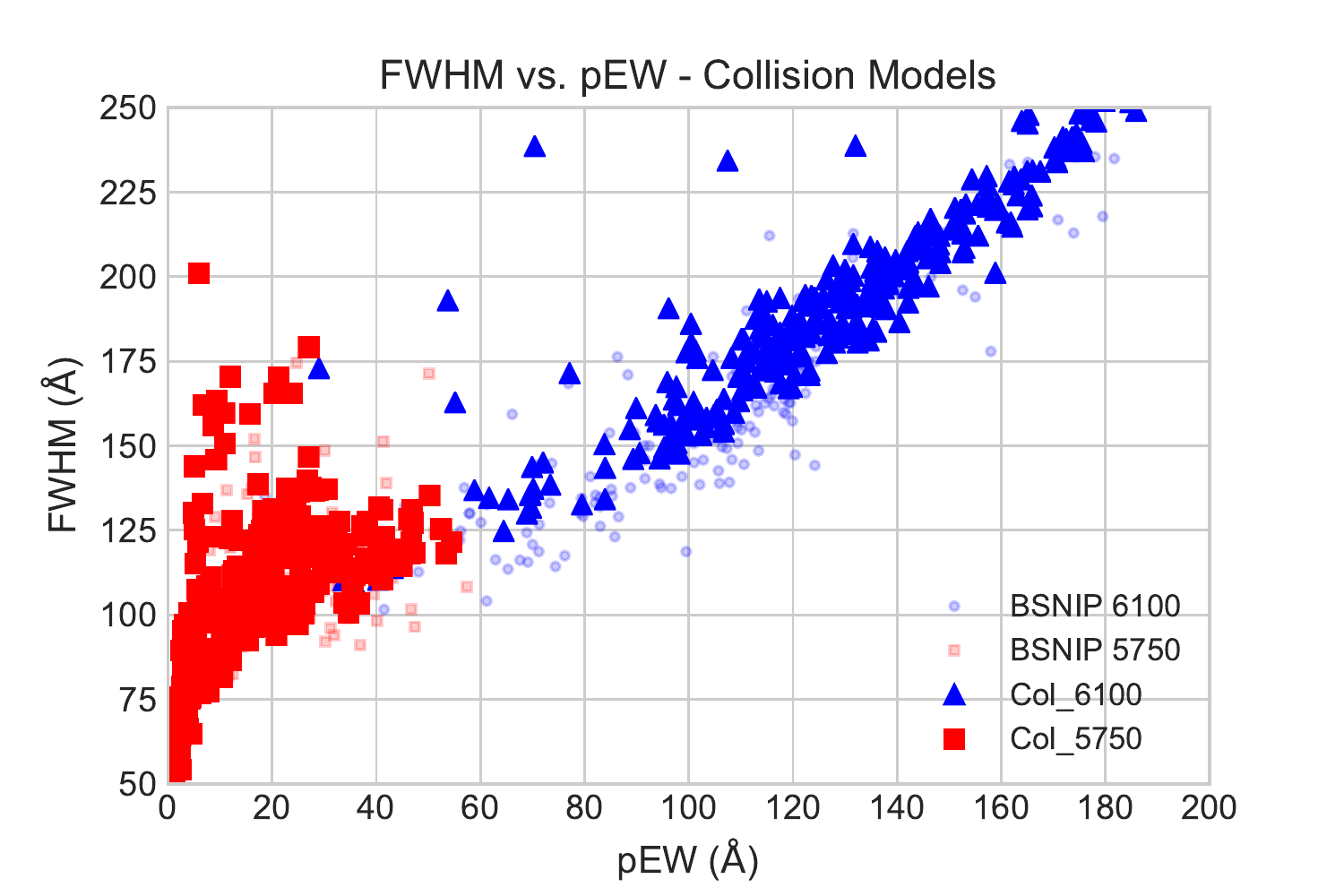}
	
	\caption{Relations between feature parameters in measured BSNIP data compared with simulated models. \textbf{Left:} Fractional depth vs. pEW for $6100\angstrom$ and $5750\angstrom$ features (blue and red dots). \textbf{Right:} Same for FWHM vs. pEW. See also \S\ref{sec:toy_model}.}
	\label{fig:simulated}
\end{figure*}

The simple toy model presented in \S\ref{sec:toy_model} and shown in Fig.~\ref{fig:feature_formation} reproduces a crude approximation of the observed relations between feature parameters. It is interesting to explore how the various simulated models fare in comparison to the observations. Fig.~\ref{fig:simulated} shows results of all models discussed in this paper compared with observed data from BSNIP. It seems that all models, even the most basic synthetic models described in \S\ref{sec:synthetic}, reproduce the observed relations between feature parameters quite well. 

%%%%%%%%%%%%%%%%%%%%%%%%%%%%%%%%%%%%%%%%%%%%%%%%%%

%%%%%%%%%%%%%%%%%%%%%%%%%%%%%%%%%%%%%%%%%%%%%%%%%%
%%%%%%%%%%%%%%%%%%%%%%%%%%%%%%%%%%%%%%%%%%%%%%%%%%
%%%%%%%%%%%%%%%%%%%%%%%%%%%%%%%%%%%%%%%%%%%%%%%%%%

\section{Head-on collision model example spectra}
\label{sec:Spectra}

\begin{figure*}
    
	\includegraphics[clip, trim=2.0cm 3.3cm 2.0cm 3.5cm, width=0.63\textwidth]{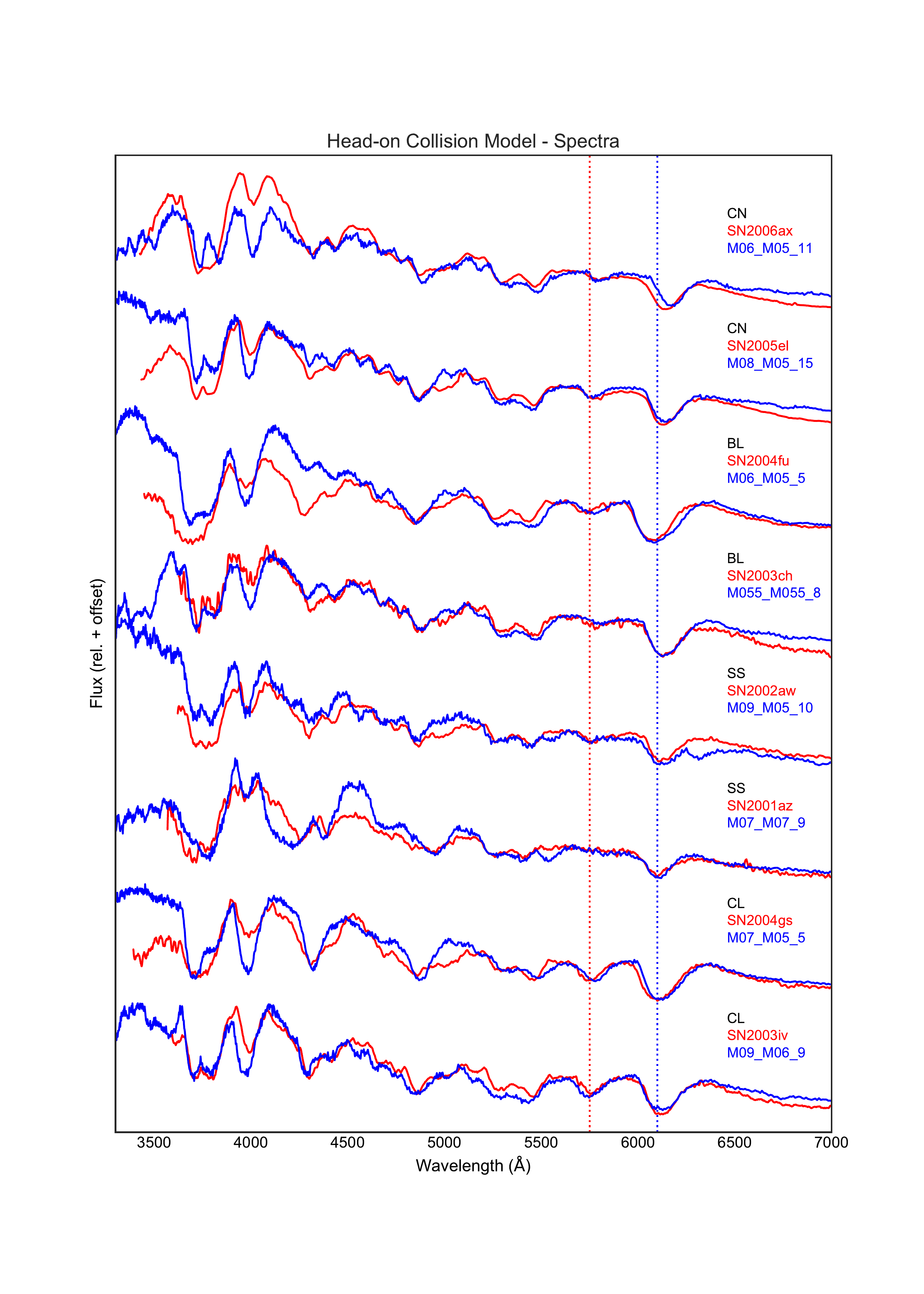} 

    \caption{Example spectra from \textsc{Tardis} simulations of head-on collision models are shown in blue. Two examples are given for each Branch type. Similar observed spectra are shown in red for comparison. The flux is scaled to match in the $6000\angstrom$ region. Observed spectra were retrieved from the WISeREP repository \citep{WiseRep}.}
	\centering
    \label{fig:Spectra}
\end{figure*}

\begin{figure*}
    \centering
	\includegraphics[clip, trim=0.8cm 0.4cm 1.0cm 0.8cm, width=\columnwidth]{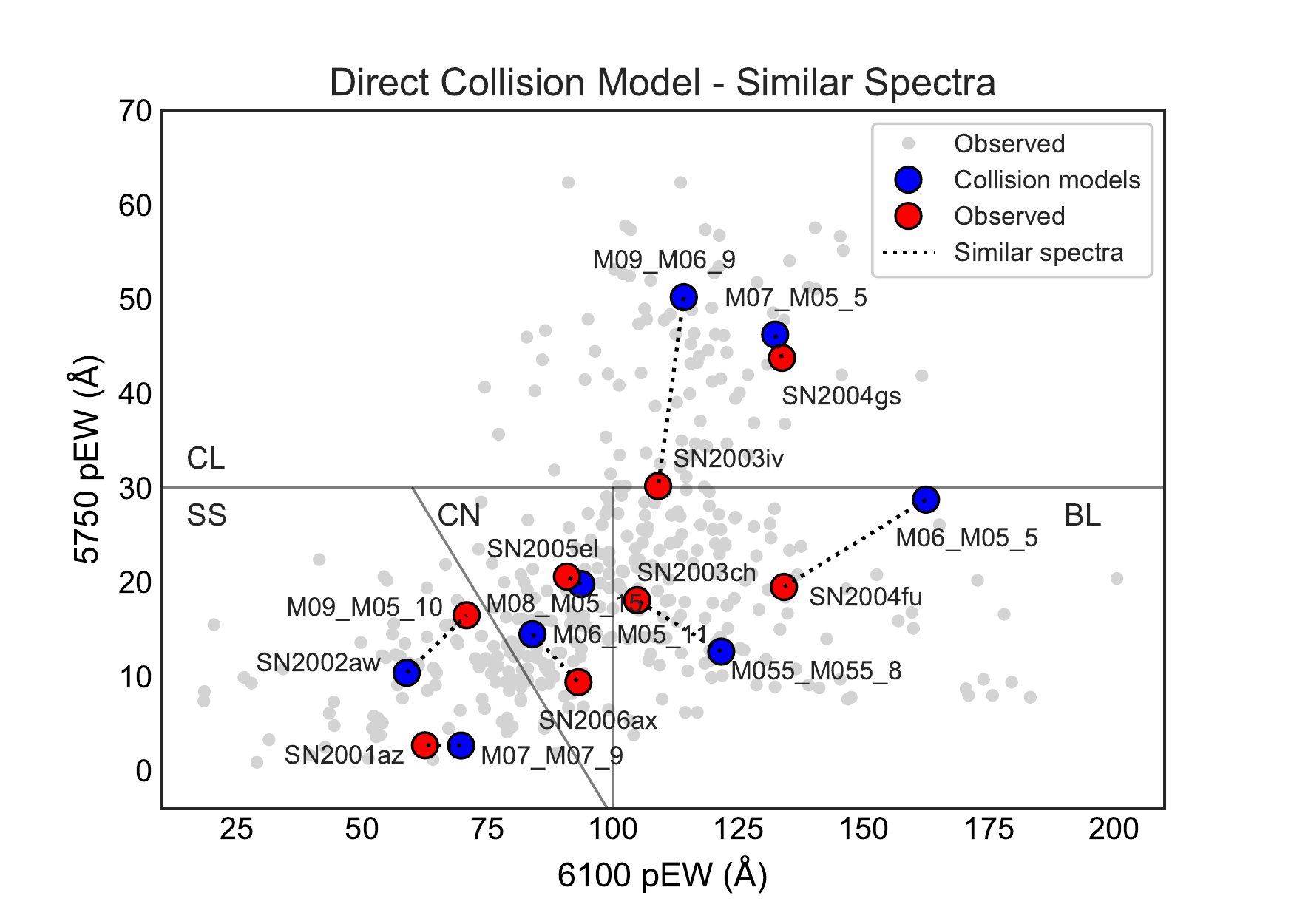}

    \caption{Branch plot showing the example head-on collision models and their corresponding similar observed events shown in Fig.~\ref{fig:Spectra}.}
    \label{fig:Neighbors}
\end{figure*}

Head-on collision model spectra are compared to observed spectra. Two models were selected for each Branch type, and an SNID-assisted \citep{SNID2007} manual search was conducted for observed spectra with similar $6100\angstrom$ and $5750\angstrom$ features. The model spectra with the closest matching observed spectra are shown in Fig.~\ref{fig:Spectra}. The location of each pair on the Branch plot is shown in Fig.~\ref{fig:Neighbors}.

%%%%%%%%%%%%%%%%%%%%%%%%%%%%%%%%%%%%%%%%%%%%%%%%%%

%%%%%%%%%%%%%%%%%%%%%%%%%%%%%%%%%%%%%%%%%%%%%%%%%%

%%%%%%%%%%%%%%%%%%%%%%%%%%%%%%%%%%%%%%%%%%%%%%%%%%
%%%%%%%%%%%%%%%%%%%%%%%%%%%%%%%%%%%%%%%%%%%%%%%%%%

\section{Synthetic models with varying Si density profile}
\label{sec:synthetic_Si}

\begin{figure*}
	\centering
	\includegraphics[width=\columnwidth]{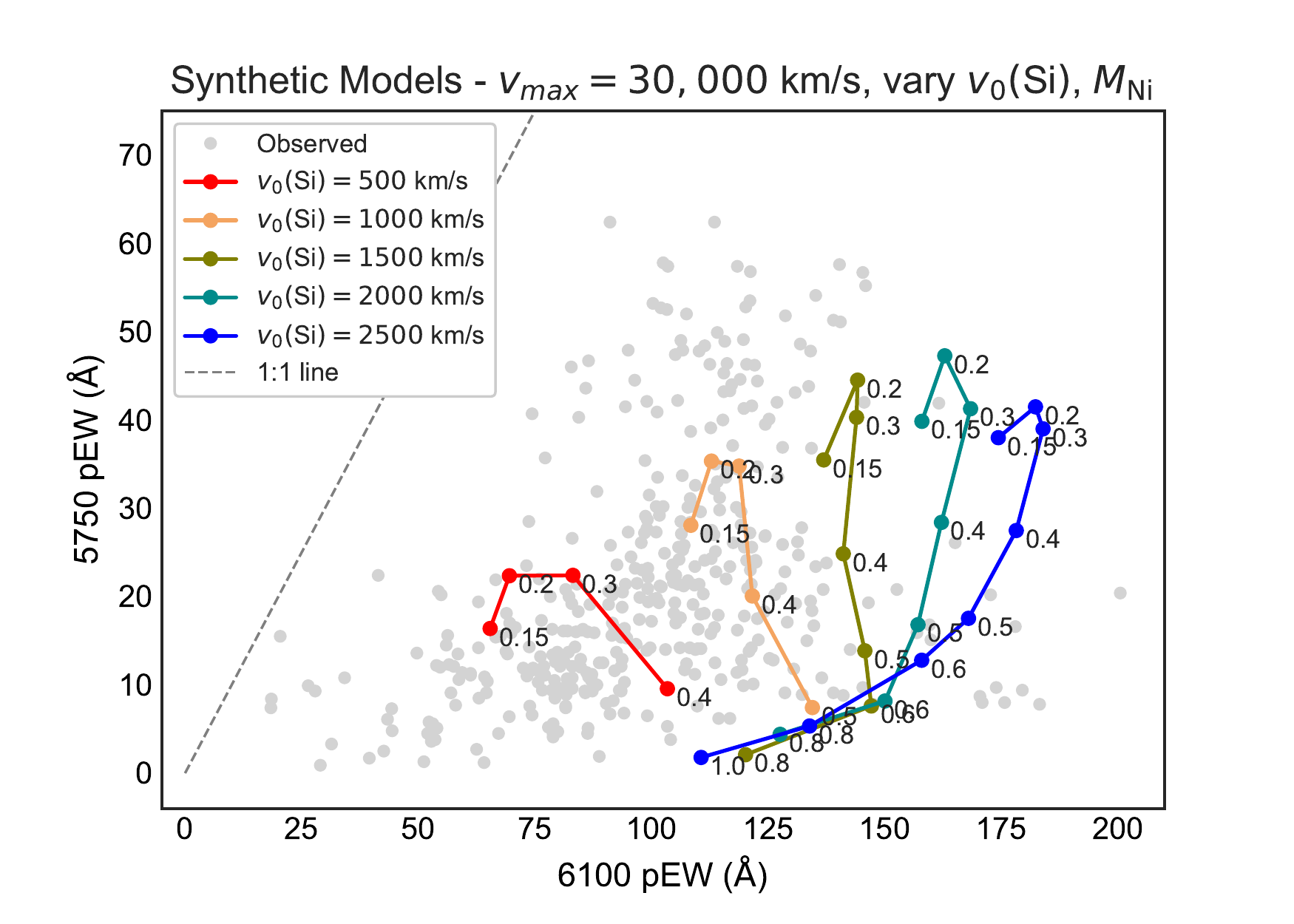}
	~ 
	\includegraphics[width=\columnwidth]{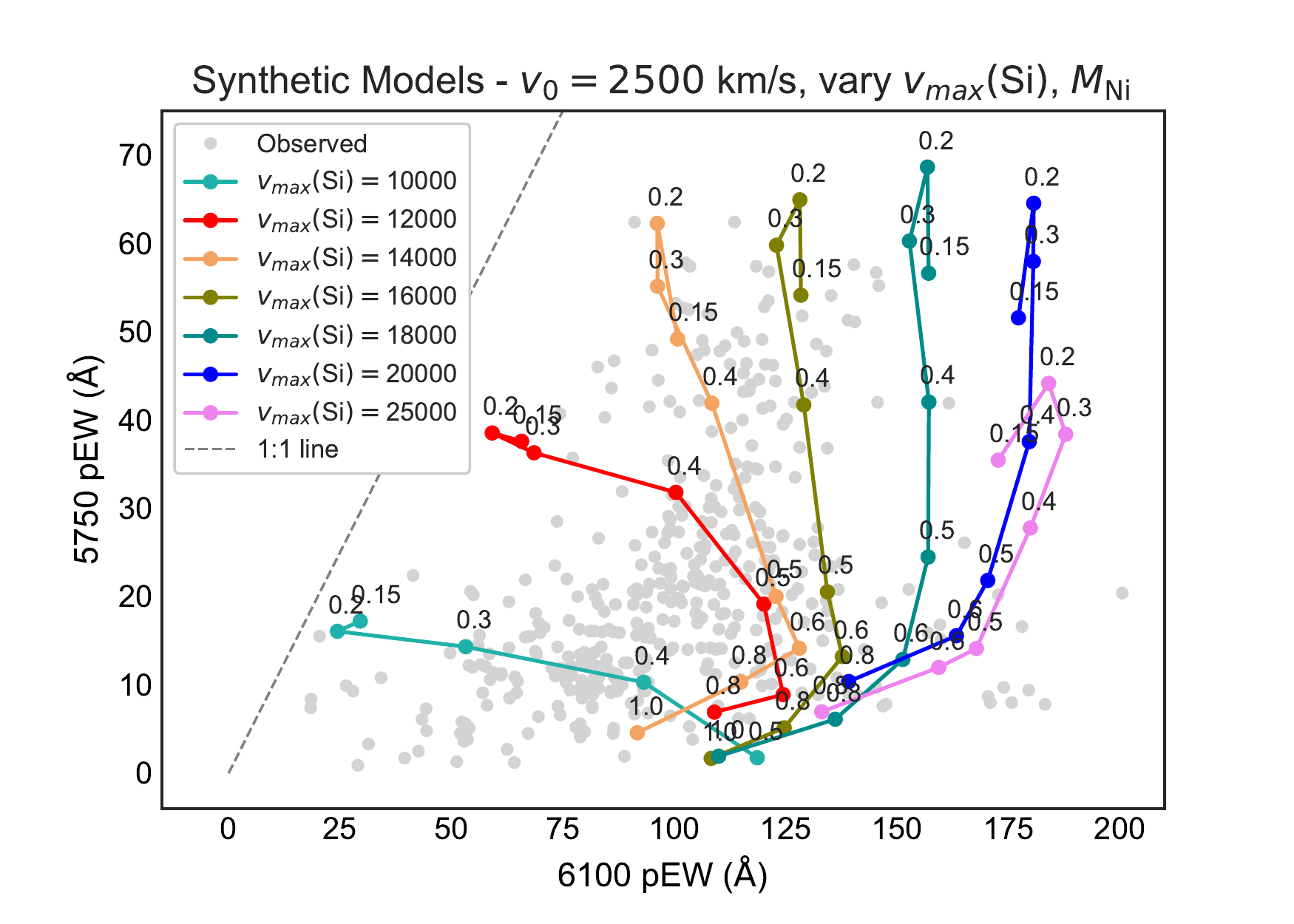}
	\caption{Simulated synthetic models with exponential density profiles of the form $\rho ({v})=\rho_0 e^{-v/v_0}$, $\rho=10^{-13}$\gcm~at $V_{\rm ph}=9,000$~km/s at 18 days, overlaid on observed CfA, CSP and BSNIP data. In all models the total density profile remains constant (with $v_0=2,500$~km/s and $v_{\max}=30,000$~km/s) and only the Si density profile is varied. Models with null or merging features have been omitted. $M(^{56}$Ni$)$ is stated next to each point in units of $\textup{M}_\odot$. $L$ is computed from $M(^{56}$Ni$)$ using Arnett's rule $L_{\max}=2.0\times10^{43}\times[M(^{56}$Ni$)/\textup{M}_\odot]$~erg/s. \textbf{Left:} Each line represents a constant Si density e-folding velocity $v_0$ with varying luminosity. \textbf{Right:} Each line represents a constant maximum Si density velocity cutoff $v_{\max}$ with varying luminosity.}
	
	\label{fig:Branch_Si}
\end{figure*}

Fig.~\ref{fig:Branch} shows plots resulting from synthetic exponentially declining models with total density: $\rho ({v})=\rho_0 e^{-{v/v_0}}$ up to a maximal velocity $v_{\max}$ (see \S\ref{sec:synthetic}). Here we show results for synthetic models in which the total density profile remains constant (with $v_0=2,500$~km/s and $v_{\max}=30,000$~km/s), while only the Si density profile is varied (either exponentially, varying $v_0$ or with a cutoff, varying $v_{\max}$). As can be seen in Fig.~\ref{fig:Branch_Si}, the results are qualitatively very similar to Fig.~\ref{fig:Branch}.

%%%%%%%%%%%%%%%%%%%%%%%%%%%%%%%%%%%%%%%%%%%%%%%%%%

%%%%%%%%%%%%%%%%%%%%%%%%%%%%%%%%%%%%%%%%%%%%%%%%%%
%%%%%%%%%%%%%%%%%%%%%%%%%%%%%%%%%%%%%%%%%%%%%%%%%%
%%%%%%%%%%%%%%%%%%%%%%%%%%%%%%%%%%%%%%%%%%%%%%%%%%

% Don't change these lines
%\bsp	% typesetting comment
\label{lastpage}
\end{document}